\documentclass[aps,twocolumn,secnumarabic,nobalancelastpage,amsmath,amssymb,nofootinbib]{revtex4-2}
\usepackage{dcolumn}                             
\usepackage{bm}                                      
\usepackage{graphicx}                             
\usepackage{amsmath}                             
\usepackage{braket}			   
\usepackage{array}			   
\usepackage{float}                                    
\usepackage{physics}                                
\usepackage[titletoc]{appendix}               

\usepackage[brazilian,british]{babel}      
\usepackage{subfigure}  		            
\usepackage[T1]{fontenc}                       
\usepackage{ae}                                      
\usepackage[utf8]{inputenc}                   

\usepackage{tikz}                                     
\usetikzlibrary{external}                           
\tikzset{external/system call={pdflatex \tikzexternalcheckshellescape -halt-on-error
-interaction=batchmode -jobname "\image" "\texsource" && 
pdftops -eps "\image.pdf"}}

\usepackage{gensymb}                           

\usepackage{booktabs}                            
\usepackage{caption}

\begin{document}

\title{The cumulant Green's functions method for the Hubbard model}


\author{R. N. Lira$^{1}$}
\author{P. S. Riseborough$^2$}
\author{J. Silva-Valencia$^3$}
\author{M. S. Figueira$^1$}
\email[corresponding author: ]{figueira7255@gmail.com}
\affiliation{Instituto de F\'{i}sica, Universidade Federal Fluminense, Av. Litor\^anea s/N, 24210-340, Niter\'oi, RJ, Brazil}
\affiliation{$^{2}$Department of Physics, Temple University, Philadelphia, PA, USA}
\affiliation{$^{3}$Departamento de F\'{\i}sica, Universidad Nacional de Colombia, A. A. 5997 Bogot\'a, Colombia.}

\begin{abstract}

We study the single-band Hubbard model under the action of an external magnetic field using the cumulant Green's functions method (CGFM). The starting point of the method is to diagonalize a cluster containing $N$ correlated sites (``seed") and employ the cumulants calculated from the cluster solution to obtain the full Green's functions for the lattice. All calculations are done directly, and no self-consistent process is needed. We benchmark the one-dimensional results for the gap, the ground-state energy, and the double occupancy obtained from the CGFM against the corresponding exact results of the thermodynamic Bethe ansatz and the quantum transfer matrix methods. The results for the CGFM tend systematically to the exact one as the cluster size increases. The particle-hole symmetry of the density of states is fulfilled. The method can be applied to any parameter space for one, two, or three-dimensional Hubbard Hamiltonians and can also be extended to other strongly correlated models, like the Anderson Hamiltonian, the $t-J$, Kondo, and Coqblin-Schrieffer models.

We also calculate the effects of positive magnetic fields as a function of the chemical potential, and we identify a finite cluster effect (Phase VI) characterized by a partially filled band and negative magnetization ($n_{up} < n_{down}$). This phase survives for clusters containing up to $N=8$ sites but tends to disappear as the size of the cluster increases. We include a simple application to spintronics, where we used these clusters as correlated quantum dots to realize a single-electron transistor when connected to Hubbard leads. We calculate the phase diagram, including the new cluster phase, using the magnetic field and chemical potential as parameters for $N=7$ and $N=8$.

\end{abstract}


\maketitle

\section{Introduction}
\label{sec1}
The Hubbard model was proposed independently in 1963 by Gutzwiller \cite{Gutzwiller1963}, Kanamori \cite{Kanamori1963}, and Hubbard \cite{HubbardI}. Hubbard worked hard to understand the model and published a series of six papers in the period $[1963-1967]$\cite{HubbardI,HubbardII,HubbardIII,HubbardIV,HubbardV,HubbardVI}, where he developed different approaches for solving it. The Hubbard Hamiltonian is the simplest interacting particle model in a lattice: it extends the tight-binding model, accounting for the electron-electron correlation (U) between electrons on the same site, not considering the effects of non-local correlations, multiple orbitals, or higher-order hoppings. It was originally developed to describe the properties of narrow partially filled $d$ band in transition metals. It has been shown that the model describes the relevant collective characteristics of these materials, namely itinerant magnetism, and metal-insulating transition. For a pertinent review see the references \cite{Essler2005,Arovas2022,Qin2022}.

The one-dimensional (1D) Hubbard model was solved exactly in a seminal paper by E. H. Lieb and F. Y. Wu \cite{LiebWu1968}, employing the technique of the Bethe ansatz \cite{Bethe1931}. They showed that it could reduce the Hamiltonian spectral problem to a set of algebraic equations. They calculated analytically the ground-state energy demonstrating that at half-filling, the model suffers a Mott metal-insulator transition \cite{Mott49, HubbardIII} at zero temperature $(T=0)$ and local critical electron correlation $U_{c}=0$ \cite{Schlottmann1995}. A complete and didactical discussion of the development of the subject can be found in the book \cite{Essler2005}.

The following fundamental advance in the Bethe ansatz formulation was attained by Takahashi, who employed a particular classification of the Lieb-Wu solutions in terms of a ``string hypothesis'' \cite{Essler2005}. He derived an infinite set of non-linear integral equations at finite temperatures and calculated the Gibbs free energy \cite{Takahashi1972, Takahashi1974, Takahashi1999}. Those integral equations are known as thermodynamic Bethe ansatz or TBA equations. They are fundamental to the study of low-temperature properties of the model but challenging to implement numerically. One step further was attained by employing a different route than TBA equations, by the development of the quantum transfer matrix (QTM) method \cite{Juttner1998}. The calculation of the properties of the 1D Hubbard model has been addressed in recent years employing different approaches associated with TBA or QTM methods \cite{Takahashi2002, Campo2015, Angela2020, Sacramento2021}. 

Another class of approximations useful in some circumstances includes the mean-field Hartree-Fock, the random phase approximation (RPA) \cite{Vignale89}, and the configurational interaction technique (CI) \cite{Berciu2000}. The latter is based on a linear combination of Hartree-Fock wave functions to restore some broken symmetries of the mean-field approach and recover some features of the exact 1D Bethe ansatz solution.

The Hubbard model has been the subject of a tremendous revival of interest in the eighties after discovering high temperature (high-$T_c$) cuprate superconductors. It has been considered the most promising model to explain strong correlations. Numerical simulations on the two-dimensional (2D) Hubbard model show regions on the parameter space that exhibits $\textit{d}-$wave superconductivity, antiferromagnetic correlations, stripes, pseudogaps, Fermi liquid, and bad metallic behavior \cite{Qin2022}. However, the connection of these phases with real high-$T_c$ superconductors is not direct.

A new interest in Hubbard model physics came from the fast and efficient experiments of ultracold atoms in optical lattices after achieving the Bose-Einstein condensation. This research area has defined an ideal platform to verify and explore new physics associated with correlated electronic systems~\cite{Esslinger-AR10, IBloch-NP12, Gross-S17}. The confinement of fermionic atoms in optical lattices allows the observation of the Mott metal-insulator transition, antiferromagnetic correlations, and spin-charge separation in one-dimensional systems with hundreds of lattice sites~\cite{Jordens-N08, Greif-Science16, Cheuk-Science16, Brown-Science17, Senaratne-ArXiv21}. Recently, the level of control and flexibility (geometry-lattice) was improved in an eight-site Fermi-Hubbard chain near half-filling achieved with lithium-6 atoms in an optical tweezer array, which allows one more motivation for the present study~\cite{Spar-ArXiv21}.

The present work belongs to a broad class of exact diagonalization (ED) methods. That generally starts from the diagonalization of a finite number of correlated sites that constitutes a cluster and employs an embedding process to reconstruct the total Hamiltonian \cite{Enrique2008}. One example of the ED method is the variational cluster approach (VCA) \cite{Potthoff2008, Seki2018}. Here, we developed the cumulant Green's functions method (CGFM) for the single-band Hubbard model in the presence of an external magnetic field. The general formalism of the cumulant expansions of the periodic Anderson model \cite{Figueira94}, as outlined here, has been previously applied by one of the authors to treat the impurity Anderson model \cite{Lobo2010} and a detailed review can be found in the arXiv repository \cite{Foglio2010}. Still, it can be generalized to the Anderson or Hubbard lattice models and variants like the $t-J$, Kondo, and Coqblin-Schrieffer models.

This work has the following structure: In section \ref{sec2}, the Hamiltonian of the model, including the presence of an external magnetic field, is introduced, and a brief discussion of its physical meaning is made. In section \ref{sec3}, we introduce the basic ideas of the cumulant approach in four steps: 1. choice of a cluster of correlated sites to be solved exactly employing ED methods; 2. using the Lehmann representation, we calculate all the atomic Green's functions, associated to the possible transitions inside the cluster of correlated sites; 3. Employing these atomic Green's functions, we obtain the atomic cumulants that will be used as an approximation to 4. calculate the lattice Green's functions. In section \ref{sec4}, we present some analytical results of the one-dimensional Hubbard model that we will employ as a benchmark for the method. In section \ref{sec5}, we benchmark the results obtained with the exact available results for the single-particle gap and the ground-state energy; we also calculate the density of states and the occupation numbers. In section \ref{sec6}, we present a discussion of the magnetic field effects and the magnetic phase diagram as a function of the chemical potential. In section \ref{sec7}, we include a simple application to spintronics, where we used these clusters as correlated quantum dots to realize a single-electron transistor when connected to Hubbard leads. Finally, in section \ref{sec8} we discuss the conclusions and perspectives of the method.

\section{The Hubbard model}
\label{sec2}

The single-band Hubbard model \cite{HubbardI} is the simplest many-body Hamiltonian that allows a relevant description of the competition between two opposite mechanisms in correlated electronic systems: the kinetic energy term that describes electrons moving from site to site in the crystal lattice, which leads to its delocalization and thus favoring metallic behavior. On the other hand, the local electronic correlation favors the localization of electrons in atomic sites, favoring the Mott transition and magnetic ordering. The single-band Hubbard model in the presence of an arbitrary magnetic field \cite{Frahm_2008,Angela2020,Sacramento2021} is given by

\begin{equation}
H=H_{0}+H_{1} ,
\label{Hubbard_model}
\end{equation}
where $H_{0}$ represents the unperturbed local terms
\begin{equation}
H_{0}=\sum_{i}\left[\epsilon_{0}({n}_{i\uparrow}+{n}_{i\downarrow})-
h_{i}m_{i}\right]+
\frac{U}{2}\sum_{i\sigma} n_{i\sigma} n_{i\bar{\sigma}} ,
\label{Ho}
\end{equation}
and the perturbation $H_{1}$ is the kinetic energy
\begin{equation}
H_{1}= -\sum_{i\neq j, \sigma}t_{ij}{c}^{\dagger}_{i\sigma}{c}_{j\sigma} .
\label{H1}
\end{equation}

The operators ${c_{i,\sigma}}^{\dagger}$ and ${c_{i\sigma}}$ represent the creation and annihilation of electrons, respectively, and $n_{i\sigma}={c_{i\sigma}}^{\dagger}{c_{i\sigma}}$ is the electron number operator. The first term of the unperturbed local Hamiltonian, 
$H_{0}$, represents the local energy $E_{0}$ of the electrons subtracted from the chemical potential $\mu$, ($\epsilon_{0}=E_{0}-\mu$), here assumed site-independent; the second term is the magnetization defined by $m_{i}={n}_{i,\uparrow}-{n}_{i,\downarrow}$, with $h_{i}$ being the site spin-dependent external magnetic field, and the last term represents the local electronic correlation term, characterized by the parameter $U$, which favors the localization of electrons on the same site. The correlation energy is responsible for the Mott transition exhibited by this Hamiltonian. In the cumulant expansion of the Hubbard model \cite{Metzner91} the kinetic energy term, $H_{1}$, is considered  the perturbation, where $(-t_{ij})$ corresponds to the electron transfer integral between the $i$ and $j$ sites of the crystal lattice. In the Hubbard model, each site has only one orbital that can either be unoccupied or occupied by no electron ($\ket{0}$), a spin-up electron ($\ket{\uparrow}$), or a spin down electron ($\ket{\downarrow} $, or by two electrons of opposing spin 
($\ket{\uparrow\downarrow}=\ket{d}$). 

\section{The cumulant Green's functions method}
\label{sec3}

The cumulant expansion of the Hubbard model was introduced by Hubbard \cite{HubbardV, HubbardVI} and applied to the infinite dimension limit by Metzner \cite{Metzner91}. He considered the local terms of Eq. \ref{Ho} as the unperturbed Hamiltonian and the kinetic energy, Eq. \ref{H1}, as a perturbation. The perturbation expansion was set up at Matsubara finite temperature $T$ representation, employing the grand-canonical ensemble.

The single-particle temperature dependent Green's functions are defined by
\begin{equation}
\label{eq:123456789123}
G_{ij\sigma}(\tau)=-\Braket{\mathcal{T}[c_{i\sigma}(\tau), c^{\dagger}_{j \sigma}(0)]},
\end{equation}
where $\mathcal{T}$ represents the temporal ordering operator for fermions and the ``time" $ \tau $ is defined in the interval $ [- \beta, \beta] $, where $ \beta = 1 / k_{B}T $, with $k_{B}$ being the Boltzmann constant.

The Green's functions diagrammatic expansion for the single Hubbard model \cite{HubbardV, HubbardVI, Metzner91, Gusmao95} can be written in terms of Feynman diagrams. The relevant diagrams for a bipartite lattice are represented up to the fourth-order in reference \cite {Metzner91}. The analysis of these diagrams shows that they are formed by irreducible parts connected by hopping lines (irreducible parts are those diagrams that cannot be divided into two pieces by cutting a single hopping line). Due to this structure, on the limit of infinite dimension, the entire perturbative series can be formally added up in the temperature Matsubara representation, resulting in a Dyson equation \cite{Metzner91}
\begin{equation}
\label{eq:55879750}
G_{\pmb{k}\sigma}(i\omega_n)=M_{\sigma}(i\omega_n)+M_{\sigma}(i\omega_n)\epsilon_{\pmb{k}}G_{\pmb{k}\sigma}(i\omega_n),
\end{equation}
where $\omega_n=(2n+1)\pi k_{B}T$ with $n=\pm 1, \pm 2,...$ correspond to Matsubara frequencies along the imaginary axis, and $\epsilon_{\pmb{k}}$ is the dispersion relation.
$M_{\sigma}(i\omega_n)$ represents the irreducible cumulants corresponding to the single-particle Green's functions (\ref{eq:123456789123}). In the infinite dimension limit, these cumulants do not depend on the wave vector $\pmb{k}$; they only depend on the Matsubara frequencies, which implies a huge simplification in the calculations, and the formal solution of (\ref {eq:55879750}) can be written as:
\begin{equation}
\label{eq:226547779}
G_{\pmb{k}\sigma}(i\omega_n)=\frac{M_{\sigma}(i\omega_n)}{1-\epsilon_{\pmb{k}}M_{\sigma}(i\omega_n)}=\frac{1}{M_{\sigma}^{-1}(i\omega_n)-\epsilon_{\pmb{k}}},
\end{equation}
or, in terms of the self-energy $\Sigma$, as commonly used in the Dynamical Mean-Field Theory (DMFT),
\begin{equation}
\label{eq:882}
M_{\sigma}^{-1}(i\omega_n)=i \omega_n + \mu - \Sigma \left(\pmb{k}\sigma,i\omega_n\right) .
\end{equation}

\begin{figure}[t]
\centering
\includegraphics[clip,width=0.45\textwidth,angle=0.]{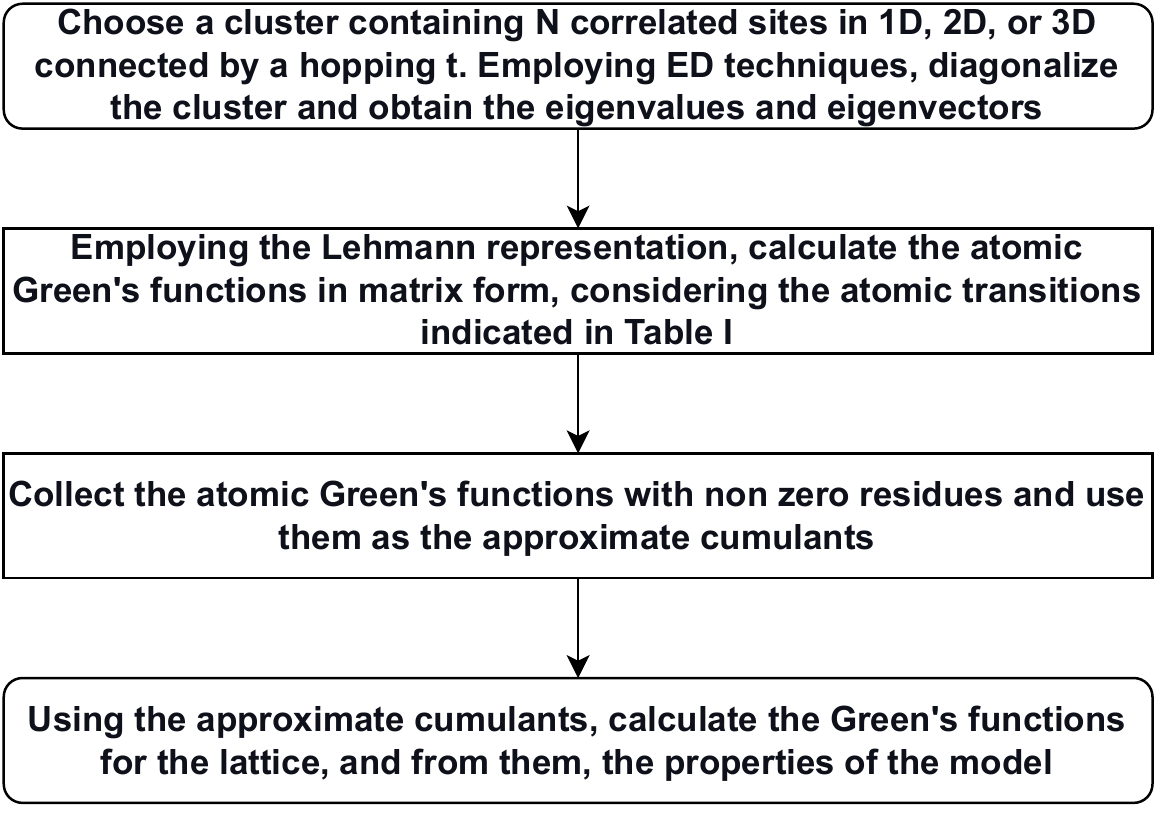}
\caption{Flowchart of the calculation steps of the CGFM.}
\label{fig:diagram}
\end{figure}
In this work, we only consider the one-dimensional version of the Hubbard model, and we will benchmark the results obtained with the exact available solutions \cite{LiebWu1968, Takahashi2002, LiebWu2003, Campo2015, Angela2020, Sacramento2021}. In the first step of the calculation, we employ ED techniques to calculate the eigenvalues and the eigenvectors of a linear cluster of $N$ Hubbard correlated sites. Due to computational limitations, we diagonalized matrices until $N=9$ sites. However, we showed that the results obtained with $N=7,8$, compared to the exact available results, produce excellent approximations to calculate the gap in the density of states,  the ground state energy (GSE), and the occupation numbers.

\begin{figure}[t]
\centering
\includegraphics[clip,width=0.45\textwidth,angle=0.]{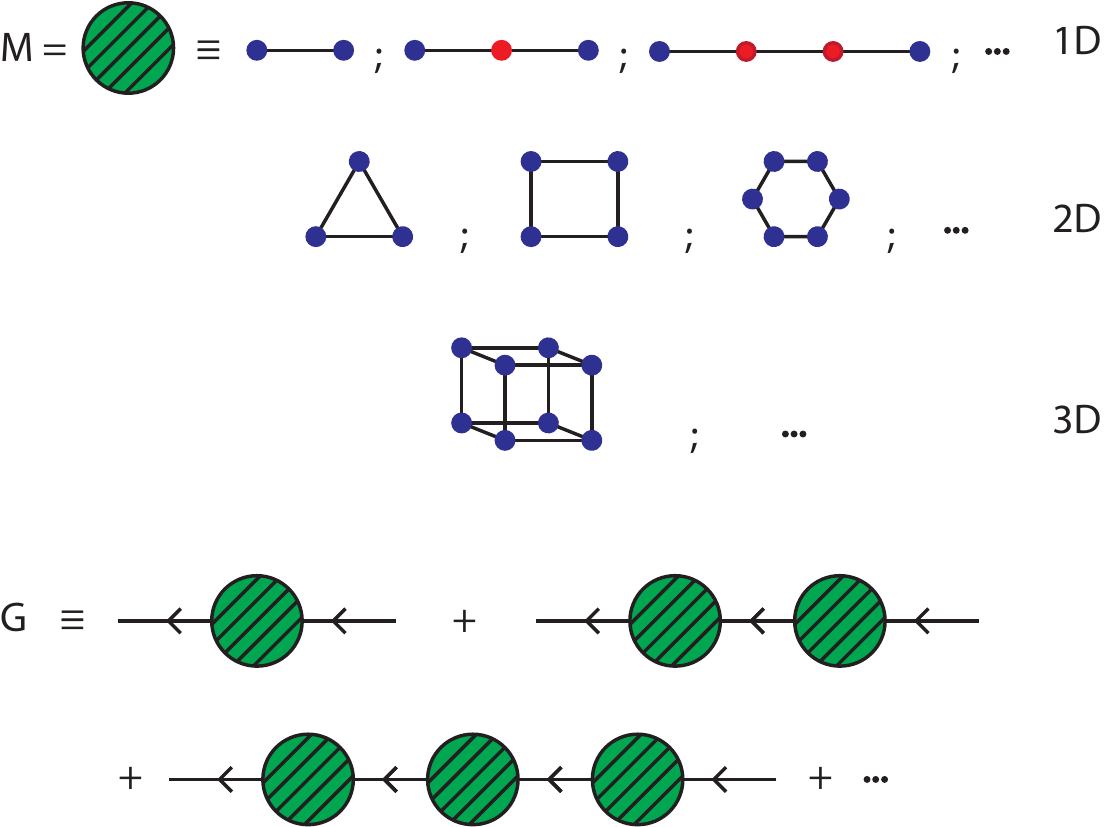}
\caption{Schematic of the type of the clusters employed as a ``seed" to generate 1D, 2D, and three-dimension (3D) cumulants, $M$ and from them the Dyson equation to calculate the lattice Green's functions $G$ of the method.}
\label{seed}
\end{figure}

Fig. \ref{seed} presents a schematic representation of the cluster employed in the calculations. In the linear chains, the different colors indicate non-equivalent correlated sites. In this case, we should perform an average of the cumulants of those sites. However, when all the sites are equivalent, as represented in the 2D and 3D examples of Fig. \ref{seed} no averages need to be performed.

The first step of the method is to choose a cluster of atoms (``seed") to calculate the atomic cumulants. In our earlier work, we employed the exact solution of the Anderson dimer \cite{ Lobo2010}. However, we will utilize the ED solution of 1D correlated site clusters larger than two for the one-dimensional Hubbard Hamiltonian. In the ED cluster calculations we used the hopping $t=1$ as the energy unit; the local energy $E_{0}$ of the electrons subtracted from the chemical potential, in the presence of the magnetic field $h$ is given by:  $\epsilon_{0} = E_{0}-\mu_\sigma$, with $\mu_\uparrow=\mu+h$, and $\mu_\downarrow=\mu-h$. It represents an important technical programming detail because we define the energies and the effects of the particle filling through the chemical potential and magnetic fields on the cluster eigenenergies. We do not need to consider those effects again during the embedding process of the cluster inside the lattice. From the eigenvalues and eigenvectors obtained, we calculate the corresponding atomic Green's functions employing the Lehmann representation and the atomic cumulants $M$ used as approximate ``seeds'' to calculate the full Green's functions $G$ of the original lattice Hamiltonian. However, the method is sufficiently general, allowing the choice of more sophisticated ``seeds" than 1D as schematically represented in Fig. \ref{seed}. We can also use ``seeds" of 2D or 3D shapes to take into account the geometry of the lattice and the richness of interactions present in 2D and 3D systems. 

The second step of the method is to use the eigenvalues, and the eigenvectors obtained earlier to calculate the atomic cluster Green's functions. This calculation represents an additional difficulty in traditional ED calculations, but it allows to take into account in a controllable way all the atomic transitions inside the cluster of correlated sites used as a ``seed" to generate the approximate cumulants. Employing the spectral representation (Lehmann representation), \cite {Zubarev1960} we obtain
\begin{equation}
\label{eq:4}
\begin{aligned}
g^{at}_{\sigma}(i\omega_{s})= {} &-e^{\beta\Omega}\sum_{n,r,r^{\prime}}\frac{\exp(-\beta\varepsilon_{n-1,r})+\exp(-\beta\varepsilon_{n,r^{\prime}})} {i\omega_{s}+\varepsilon_{n-1,r}-\varepsilon_{n,r^{\prime}}} \\
& \times \Bra{n-1,r}c_{i\sigma}\Ket{n,r^{\prime}} \Bra{n,r^{\prime}}c_{i\sigma}^{\dagger}\Ket{n-1,r} ,
\end{aligned}
\end{equation}
where $\Omega$ is the grand canonical potential, with $\beta=1/k_{B}T$, and the eigenvectors $\ket{n,r}$ and eigenvalues $\varepsilon_{n,r}$ correspond to the complete solution of the atomic cluster Hamiltonian. The atomic Green's function can be rewritten as:
\begin{equation}  
\label{eq:5}
g^{at}_{\sigma}(i\omega_{s}) = e^{\beta\Omega} \sum_{i} \frac{r_{i,\sigma}}{i\omega_{s}-u_{i,\sigma}} ,
\end{equation}
where $u_{i,\sigma}$ are the poles and $ r_{i,\sigma} $  the residues of the atomic Green's functions, respectively.

We should calculate the atomic Green's functions in matrix form, considering the electron spin destruction (creation) in the allowed atomic transitions. Here, $ n $ to $ n \pm 1 $ electrons, indicates the total number of electrons of the considered state as indicated in Table \ref{tab:3}. This procedure is a central point of the method, differentiating it from other ED approaches like the VCA \cite{Potthoff2008, Enrique2008, Seki2018}. The focus of the calculation is on the possible atomic transitions within the atomic cluster.
\begin{table}
\centering
{\setlength{\extrarowheight}{3.0pt}
{\renewcommand{\arraystretch}{1.0}
\begin{tabular}{| c | c | c | c | c | }
\hline
$I_{x}$ & $1$ & $2$ & $3$ & $4$\\
\hline
$\alpha=(b,a)$ & $(0,\uparrow)$ & $(0,\downarrow)$ & $(\downarrow,d)$ & $(\uparrow,d)$\\\hline\hline
$g^{at}$& $g_{11}$ & $g_{33}$ & $g_{13}$ & $g_{31}$ \\
\hline
 $ I_{x}=1,3 $   & $(0,\uparrow)$ & $(\downarrow,d)$ & $(0,\uparrow)$ and 
$(\downarrow,d)$ & $(\downarrow,d)$ and $(0,\uparrow)$ \\
\hline\hline
$g^{at}$ & $g_{22}$ & $g_{44}$ & $g_{24}$ & $g_{42}$ \\
\hline
$I_{x}=2,4$   & $(0,\downarrow)$ & $(\uparrow,d)$ & $(0,\downarrow)$ and 
$(\uparrow,d)$ & $(\uparrow,d)$ and $(0,\downarrow)$ \\
\hline
\end{tabular}}}
\caption{(Above) Representation of the possible transitions present in the Hubbard Hamiltonian, where in $\alpha=(b,a)$, $a$ represents the initial state and $b$ the final state. $I_{x}=1,3$ destroy one electron with spin up and $I_{x}=2,4$ destroy one electron with spin down. We use $\sigma=\uparrow$ and $\sigma=\downarrow$ to represent the up and down spins, respectively. The double occupation state is represented by the label $d$. (Below) Atomic Green's functions associated to the  processes $I_{x}=1,3$ and $I_{x}=2,4$.}
\label{tab:3}
\end{table}

We define the atomic Green's functions $ g_ {11} $, $ g_ {33} $, $ g_ {13} $ and $ g_ {31} $ associated with transitions that destroy a spin-up electron  and the functions $ g_ {22} $, $ g_ {44} $, $ g_ {24} $ and $ g_ {42} $, associated with transitions that destroy a spin down electron, as detailed in table \ref {tab:3} (the superscript ``at" is not used on the matrix components for simplicity). Also, according to table \ref {tab:3}, it can be seen that functions $ g_ {11} $ and $ g_ {22}$ are associated with states that initially contain a single electron, whereas $ g_ {33} $ and $ g_ {44}$ are associated with states that initially contain two electrons and $ g_ {13} $, $ g_ {31}$, $ g_ {24} $ and $ g_ {42}$ are the crossed GFs and are associated to the simultaneous destruction of electrons in states of single and double occupations. The atomic Green's functions associated with all of the allowed transitions within the cluster of correlated sites considered are calculated, separated, and indexed.  Thus, one can write the atomic Green's functions as
\begin{equation}
\label{eq:771}
\mathbf{g}^{\text{at}}_{\sigma}\left(i\omega\right) = \left( \begin{array}{cccc}
g_{11} & g_{13}  & 0 & 0 \\
g_{31}  & g_{33}  & 0 & 0 \\
0 & 0 & g_{22}  & g_{24}  \\
0 & 0 & g_{42}  & g_{44} 
\end{array} \right).
\end{equation}
Equation (\ref {eq:771}) presents itself in a diagonal block form because the selection rules do not allow transitions with spin inversion, and the spins up and down are disconnected.

In the third step of the method, we collect the atomic Green's functions (\ref{eq:771}) associated with the possible atomic transitions with residues different from zero and use them as the approximate atomic cumulants. They belong to the most straightforward class of cumulants that are connected by two Fermi-Dirac lines, as discussed by Hubbard in his fifth paper about cumulant expansions of the Hubbard model \cite{HubbardV}.

$$
\mathbf{m}_{\sigma}^{\text{at}}\left(i\omega\right)= 
$$
\begin{equation}
\label{cum2}
\left( \begin{array}{cccc}
m_{11} & m_{13}  & 0 & 0 \\
m_{31}  & m_{33}  & 0 & 0 \\
0 & 0 & m_{22}  & m_{24}  \\
0 & 0 & m_{42}  & m_{44} 
\end{array} \right)=\left( \begin{array}{cccc}
g_{11} & g_{13}  & 0 & 0 \\
g_{31}  & g_{33}  & 0 & 0 \\
0 & 0 & g_{22}  & g_{24}  \\
0 & 0 & g_{42}  & g_{44} 
\end{array} \right).
\end{equation}

Finally, in the fourth step of the method, we use the atomic cumulants of Eq. \ref{cum2} as  an approximation to the formally exact cumulant $ {M_ {\sigma} (i \omega_n)} $ to calculate the Green's functions for the lattice $ {G_ {\pmb{k} \sigma} (i \omega_n)} $ and from them the properties of the model: Ground state energy, gap, occupation numbers, the density of states (DOS) and other dynamical properties of the model. A flowchart of all the steps of the method is presented in Fig. \ref{fig:diagram}

Due to the nature of the method, there will always be the simple structure represented by $4$x$4$ deblocked matrix (\ref {eq:771}) for the atomic Green's functions and the atomic cumulants \ref{cum2} respectively, regardless of the size of the atomic cluster used in the calculation.  Using the same matrix form introduced in Eq. \ref{eq:771} before and carrying out the analytical continuation of the cumulant Green's functions, Eq. (\ref {eq:226547779}), to the real frequency axis, the Green's functions for the lattice become

\begin{equation}
\label{eq:797979}
\mathbf{G}_{\pmb{k}\sigma}(\omega)=\mathbf{M}_{\sigma}(\omega)\cdot \left[ \mathbf{I-\mathbf{W}_{\pmb{k}\sigma }\cdot M_{\sigma}(\omega)}\right]^{-1}.
\end{equation}

Defining the exact cumulants and the Green's functions as

\begin{equation}
\mathbf{M}_{\uparrow }(\omega)=%
\begin{pmatrix}
M_{11} & M_{13} \\ 
M_{31} & M_{33}%
\end{pmatrix}%
\hspace{20pt}\mathrm{;\hspace{20pt}}\mathbf{M}_{\downarrow }(\omega)=%
\begin{pmatrix}
M_{22} & M_{24} \\ 
M_{42} & M_{44}%
\end{pmatrix}%
,  \label{E5.97}
\end{equation}%

\begin{equation}
\mathbf{G}_{\pmb{k}\uparrow }(\omega)=%
\begin{pmatrix}
G_{11} & G_{13} \\ 
G_{31} & G_{33}%
\end{pmatrix}%
\hspace{20pt}\mathrm{;\hspace{20pt}}\mathbf{G}_{\pmb{k}\downarrow }(\omega)=%
\begin{pmatrix}
G_{22} & G_{24} \\ 
G_{42} & G_{44}%
\end{pmatrix}%
,  \label{E5.97}
\end{equation}%
one obtains the exact Green's functions $\mathbf{G}_{\pmb{k}\sigma }(\omega )$ by performing the matrix inversion in Eq. \ref{eq:797979}:
\begin{equation}
\mathbf{G}_{\pmb{k}\uparrow}(\omega)=
\begin{pmatrix}
M_{11}&M_{13} \\
M_{31}&M_{33}%
\end{pmatrix}
\left[\begin{pmatrix}
1 & 0 \\
0 & 1%
\end{pmatrix}
-\mathbf{W}_{\pmb{k}\uparrow} 
\begin{pmatrix}
M_{11} & M_{13} \\
M_{31} & M_{33}%
\end{pmatrix}
\right]^{-1} 
\end{equation}
and
\begin{equation}
\mathbf{G}_{\pmb{k}\downarrow}(\omega)=
\begin{pmatrix}
M_{22} & M_{24} \\
M_{42} & M_{44}%
\end{pmatrix}
\left[\begin{pmatrix}
1 & 0 \\
0 & 1%
\end{pmatrix}
-\mathbf{W}_{\pmb{k}\downarrow}  
\begin{pmatrix}
M_{22} & M_{24} \\
M_{42} & M_{44}%
\end{pmatrix}
\right]^{-1} 
\end{equation}
with
\begin{equation}
\mathbf{W}_{\pmb{k}\sigma }=\epsilon_{\pmb{k}}\cdot
\begin{pmatrix}
1 & -1 \\ 
-1 & 1%
\end{pmatrix}  .
\end{equation}
Performing the calculations, it follows that
\begin{equation}
\label{eq:789988}
\mathbf{G}_{\pmb{k}\uparrow}(\omega)=\frac{1}{1-\epsilon_{\pmb{k}}\Gamma_{13}}\left[\left( \begin{array}{cc}M_{11} & M_{13} \\ M_{31} & M_{33} \\ \end{array} \right) - \epsilon_{\pmb{k}}\Theta_{13}\left( \begin{array}{cc}1 & -1 \\ -1 & 1 \\ \end{array} \right)\right]
\end{equation}
and
\begin{equation}
\label{eq:799988}
\mathbf{G}_{\pmb{k}\downarrow}(\omega)=\frac{1}{1-\epsilon_{\pmb{k}}\Gamma_{24}}\left[\left( \begin{array}{cc}M_{22} & M_{24} \\ M_{42} & M_{44} \\ \end{array} \right) - \epsilon_{\pmb{k}}\Theta_{24}\left( \begin{array}{cc}1 & -1 \\ -1 & 1 \\ \end{array} \right)\right]
\end{equation}
where $\Theta_{13}=M_{11}M_{33}-M_{13}M_{31}$, $\Gamma_{13}=M_{11}+M_{13}+M_{31}+M_{33}$, $\Theta_{24}=M_{22}M_{44}-M_{24}M_{42}$ and $\Gamma_{24}=M_{22}+M_{24}+M_{42}+M_{44}$.

For simplicity, all the results of this work were calculated considering an uncorrelated  rectangular conduction band of bandwidth 2D defined by
\begin{equation} 
\label{SB1}
\rho_{0}( E_{\bf k\sigma}) =
\left\{
    \begin{array}{l}
     \frac{1}{2D} \; ,
     \mbox{ for } -D \le E_{\bf k\sigma} \le D\\
      0 \; ~~~, \mbox{ otherwise }
    \end{array}  \right. \, ,
\end{equation}
with the $U=0$ corresponding GF is given by
\begin{equation}
G^{0}_{\sigma}(\omega)=\frac{1}{2D}\ln\left(  \frac{\omega+D}{\omega-D}\right).
\label{SB2}
\end{equation}

Integrating $\mathbf{G}_{\pmb{k}\uparrow}(\omega)$ we obtain  the total
N-site Hubbard rectangular band, with the spin up  Green's function $G_{\uparrow}\left(\omega\right)$ given by
\begin{equation}
\label{eq:992}
\begin{aligned}
G_{\uparrow}\left(\omega\right)&=G_{11}\left(\omega\right)+G_{13}\left(\omega\right)+G_{31}\left(\omega\right)+G_{33}\left(\omega\right)\\ &=\frac{1}{2D}\ln\left(\frac{1+D \Gamma_{13}}{1-D\Gamma_{13}}\right),
\end{aligned}
\end{equation}
where
\begin{equation}
\label{eq:1992}
G_{11}\left(\omega\right)=\frac{\Theta_{13}}{\Gamma_{13}}+\left[M_{11}-\frac{\Theta_{13}}{\Gamma_{13}}\right]\frac{1}{2D\Gamma_{13}}\ln\left(\frac{1+D  \Gamma_{13}}{1-D\Gamma_{13}}\right),
\end{equation}
\begin{equation}
\label{eq:2992}
G_{13}\left(\omega\right)=-\frac{\Theta_{13}}{\Gamma_{13}}+\left[M_{13}+\frac{\Theta_{13}}{\Gamma_{13}}\right]\frac{1}{2D\Gamma_{13}}\ln\left(\frac{1+D  \Gamma_{13}}{1-D\Gamma_{13}}\right),
\end{equation}
\begin{equation}
\label{eq:3992}
G_{13}\left(\omega\right)=-\frac{\Theta_{13}}{\Gamma_{13}}+\left[M_{31}+\frac{\Theta_{13}}{\Gamma_{13}}\right]\frac{1}{2D\Gamma_{13}}\ln\left(\frac{1+D  \Gamma_{13}}{1-D\Gamma_{13}}\right),
\end{equation}
\begin{equation}
\label{eq:4992}
G_{33}\left(\omega\right)=\frac{\Theta_{13}}{\Gamma_{13}}+\left[M_{33}-\frac{\Theta_{13}}{\Gamma_{13}}\right]\frac{1}{2D\Gamma_{13}}\ln\left(\frac{1+D \Gamma_{13}}{1-D\Gamma_{13}}\right),
\end{equation}

The results are analogous those obtained using the total spin down Green's function $G_{\downarrow}\left(\omega\right)$
\begin{equation}
\label{eq:222}
\begin{aligned}
G_{\downarrow}\left(\omega\right)&=G_{22}\left(\omega\right)+G_{24}\left(\omega\right)+G_{42}\left(\omega\right)+G_{44}\left(\omega\right)\\ &=\frac{1}{2D}\ln\left(\frac{1+D\Gamma_{24}}{1-D\Gamma_{24}}\right),
\end{aligned}
\end{equation}

where
\begin{equation}
\label{eq:1892}
G_{22}\left(\omega\right)=\frac{\Theta_{24}}{\Gamma_{24}}+\left[M_{22}-\frac{\Theta_{24}}{\Gamma_{24}}\right]\frac{1}{2D\Gamma_{24}}\ln\left(\frac{1+D \Gamma_{24}}{1-D\Gamma_{24}}\right),
\end{equation}
\begin{equation}
\label{eq:2892}
G_{24}\left(\omega\right)=-\frac{\Theta_{24}}{\Gamma_{24}}+\left[M_{24}+\frac{\Theta_{24}}{\Gamma_{24}}\right]\frac{1}{2D\Gamma_{24}}\ln\left(\frac{1+D \Gamma_{24}}{1-D\Gamma_{24}}\right),
\end{equation}
\begin{equation}
\label{eq:3892}
G_{42}\left(\omega\right)=-\frac{\Theta_{24}}{\Gamma_{24}}+\left[M_{42}+\frac{\Theta_{24}}{\Gamma_{24}}\right]\frac{1}{2D\Gamma_{24}}\ln\left(\frac{1+D \Gamma_{24}}{1-D\Gamma_{24}}\right),
\end{equation}
\begin{equation}
\label{eq:4892}
G_{44}\left(\omega\right)=\frac{\Theta_{24}}{\Gamma_{24}}+\left[M_{44}-\frac{\Theta_{24}}{\Gamma_{24}}\right]\frac{1}{2D\Gamma_{24}}\ln\left(\frac{1+D \Gamma_{24}}{1-D\Gamma_{24}}\right).
\end{equation}

and the total Green's function is given by
\begin{equation}
G_{\sigma}(\omega)=G_{\uparrow}\left(\omega\right)+G_{\downarrow}\left(\omega\right) .
\label{Gtot}
\end{equation}

Following the standard procedure (substituting $\omega$ by $\omega + i\eta$, taking the limit as $\eta \to 0^+ $), the DOS can be written as
\begin{equation}
\label{eq:447721}
\rho_{\sigma} (\omega)=\frac{1}{\pi}\Im{G_{\sigma}(\omega)}.
\end{equation}

In principle, Eq. (\ref {eq:797979}) and all the Green's functions obtained from it are exact in the infinite dimension limit. However, as the full lattice cumulants, $\mathbf{M}_{\sigma}\left(\omega\right)$, equation \ref{E5.97}, are unknown, the atomic cumulants $\mathbf{m}_{\sigma}^{\text{at}}\left(\omega\right)$ obtained from Eq. \ref{cum2} as the solution of a cluster containing $ N $ correlated sites are used as approximations to calculate the Green's functions of the lattice: equations (\ref {eq:992}) to (\ref{Gtot}). The method shows its full potential here because we can use ED to solve an increasing cluster of Hubbard correlated sites, and from these solutions, it is possible to build better approximations that are increasingly closer to the exact solution of the problem and that satisfy the completeness relation of the Hubbard model given by
\begin{equation}
comp^{\sigma}=n^{\sigma}_{vac}+n^{\sigma}_{up}+n^{\sigma}_{down}+n^{\sigma}_{d}=1,  \label{Occ1}
\end{equation}
where $\sigma=(\uparrow;\downarrow)$, with  $\sigma=\uparrow$ representing the transitions associated with $I_{x}=1,3$, that correspond to a destruction of a spin-up electron and $\sigma=\downarrow$ those associated with $I_{x}=2,4$, which corresponds to a destruction of a spin down electron (see the Table \ref{tab:3}). The first term is the vacuum occupation number, the second and the third terms are the occupation of the spin-up and down, respectively, and the last one is the double occupation. All the different averages could be calculated employing the Green's functions $G_{11}(\omega )$ and $G_{33}(\omega )$ associated with the processes $I_{x}=1,3$, and defined by Eqs. \ref{eq:1992} and \ref{eq:4992}:
\begin{equation}
n^{\uparrow}_ {vac}=\left( \frac{1}{\pi }\right)
\int_{-\infty }^{\infty }d\omega Im(G_{11})(1-n_{F}),  \label{G001}
\end{equation}%
\begin{equation}
n^{\uparrow}_{up}=\left( \frac{1}{\pi }\right)
\int_{-\infty }^{\infty }d\omega Im(G_{11})n_{F},  \label{Gup1}
\end{equation}%
\begin{equation}
n^{\uparrow}_ {down}=\left( \frac{1}{\pi }\right)
\int_{-\infty }^{\infty }d\omega Im(G_{33})(1-n_{F}),  \label{Gdown1}
\end{equation}%
\begin{equation}
n^{\uparrow}_ {d}=\left( \frac{1}{\pi }\right)
\int_{-\infty }^{\infty }d\omega Im(G_{33})n_{F},  \label{Gdd1}
\end{equation}%
where $n_{F}(x)=1/\left[ 1+\exp (\beta x)\right]$ is the Fermi-Dirac distribution. 

Similarly, we could employ the Green's functions $G_{22}$ and $G_{44}$, and associated with the destruction of electrons processes $I_{x}=2,4$ to calculate the corresponding occupation numbers.

\begin{equation}
n^{\downarrow}_ {vac}=\left( \frac{1}{\pi }\right)
\int_{-\infty }^{\infty }d\omega Im(G_{22})(1-n_{F}),  \label{G002}
\end{equation}%
\begin{equation}
n^{\downarrow}_{down}=\left( \frac{1}{\pi }\right)
\int_{-\infty }^{\infty }d\omega Im(G_{22})n_{F},  \label{Gup2}
\end{equation}%
\begin{equation}
n^{\downarrow}_ {up}=\left( \frac{1}{\pi }\right)
\int_{-\infty }^{\infty }d\omega Im(G_{44})(1-n_{F}),  \label{Gdown2}
\end{equation}%
\begin{equation}
n^{\downarrow}_ {d}=\left( \frac{1}{\pi }\right)
\int_{-\infty }^{\infty }d\omega Im(G_{44})n_{F},  \label{Gdd2}
\end{equation}%

The completeness relation and the full occupation numbers are then
\begin{equation}
comp=comp^{\uparrow}+comp^{\downarrow}, \label{OccT}
\end{equation}
\begin{equation}
n_{vac}=n^{\uparrow}_ {vac}+n^{\downarrow}_ {vac}, \label{G00T}
\end{equation}
\begin{equation}
n_{up}=n^{\uparrow}_{up}+n^{\downarrow}_{up}, \label{GupT}
\end{equation}
\begin{equation}
n_ {down}=n^{\uparrow}_ {down}+n^{\downarrow}_ {down}, \label{GdownT}
\end{equation}
\begin{equation}
n_ {d}=n^{\uparrow}_ {d}+n^{\downarrow}_ {d}. \label{GddT}
\end{equation}

The electron density $n$ per lattice site (electron concentration or band filling) is defined according to Eqs. \ref{GupT}-\ref{GddT} as
\begin{equation}
n=\frac{N_{e}}{N_{latt}} =n_{up}+n_ {down}+2n_ {d},
\label{Density}
\end{equation}
where $N_{e}$ is the electron number and $N_{latt}$ is total number of lattice sites. The factor $2$ in front of $n_{d}$ refers to the number of electrons inside of the double occupied state. The maximum number of electrons per site in the Hubbard model is $2$, and the limiting cases $n = 0$ and $n = 1$ correspond to empty and half-filled bands, respectively.

\section{Analytical results of the one-dimensional Hubbard model}
\label{sec4}

The one-dimensional Hubbard model was solved analytically at half-filling employing the technique of the Bethe Ansatz \cite{LiebWu1968}. In this limit, the model has particle-hole symmetry, and the band is half-filled; some important analytical results were obtained for the GSE and the density of states gap, as indicated in the references \cite{LiebWu1968, Ovchinnikov70, Essler2005, Potthoff2008} and references within. The ground-state energy $E_{g}/L$ for $\epsilon_{0}=E_{0}- \mu =-U / 2 $, with $ L $ denoting the number of sites and the exact expression for the gap in the DOS, are given, respectively by

\begin{equation} 
\label{ground-state_ba}
E_{g}/L=-4t \int_{0}^{\infty}dx\frac{J_0(x)J_1(x)}{x[1+\exp(xU/2t)]},
\end{equation}
where $J_0$ and $J_1$ are Bessel functions,

\begin{equation}
\label{eq:gap_ba}
\Delta=\frac{16t^{2}}{U}\int_{1}^{\infty}dx\frac{\sqrt{x^2-1}}{\sinh(2\pi tx/U)}.
\end{equation}
There is a gap for any $ U> 0 $ value and no critical finite $ U $ for the Mott insulator-metal transition. These analytical results will be used later to benchmark the CGFM.

In general, according to reference \cite {Vignale89} and references within, the ground-state energy $ E_{g} (U) $ of the single-band Hubbard Hamiltonian is given by
\begin{equation}
E_{g}(U)=E_{g}(U=0)+\int_{0}^{U}\sum_{i}\left<n_{i\uparrow}n_{i\downarrow}\right>_{U'} dU',
\label{Vignale}
\end{equation}
where, the term $\left<n_{i\uparrow}n_{i\downarrow}\right>$ is the number of doubly occupied states, $ n_ {d} $, that can be calculated by means of the CGFM, using equations 
(\ref {Gdd1} and \ref{Gdd2}). $ E_{g}(U=0) $ is the ground-state energy of the non-interacting system (no electronic correlation) \cite{Jafari2008}
$$
E_{g}(U=0)=\sum_{|\pmb{k}|<\pi/2,\sigma}\epsilon_{\pmb{k}}n_{\pmb{k}\sigma}=
$$
\begin{equation}
\int_{-\pi/2}^{\pi/2}dk \frac{1}{2\pi}2(-2t\cos k)=-\frac{4t}{\pi} .
\end{equation}

Another significant result is the interpolative double occupation formula \cite{Baeriswyl91}
\begin{equation}
n_{d}=\frac{1+c_{1}U}{4(1+c_{2}U+c_{3}U^{2}+c_{4}U^{3})} ,
\label{Double_occ}
\end{equation}
where for 1D $c_{1}=2.445$, $c_{2}=2.581$, $c_{3}=0.090$, and $c_{4}=0.220$. This formula also works for 2D and 3D, with different coefficients \cite{Baeriswyl91}.

\section{Results and discussion}
\label{sec5}

\begin{figure}[t]
\centering
\includegraphics[clip,width=0.45\textwidth,angle=0.]{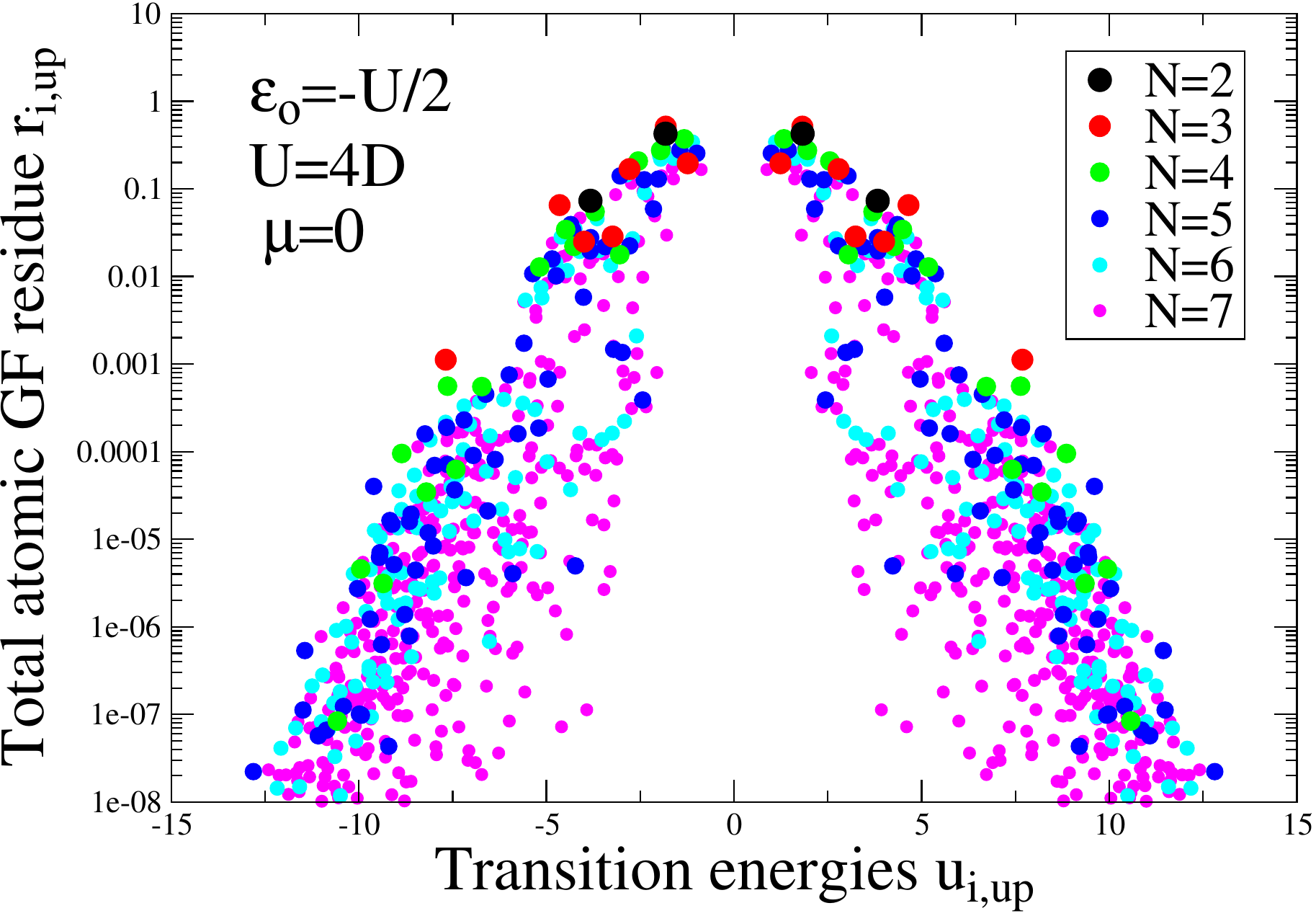}
\caption{Numerical value of the residues $r_{i,up}$ of the total atomic GF, Eq. \ref{eq:5}: $g^{at}_{up}=g_{11}+g_{13}+g_{31}+g_{33}$,  as a function of the transition energy $u_{i,up}$ for $N$ size clusters of correlated sites.}
\label{fig:weight_residues}
\end{figure}
\begin{figure}[t]
\centering
\includegraphics[clip,width=0.45\textwidth,angle=0.]{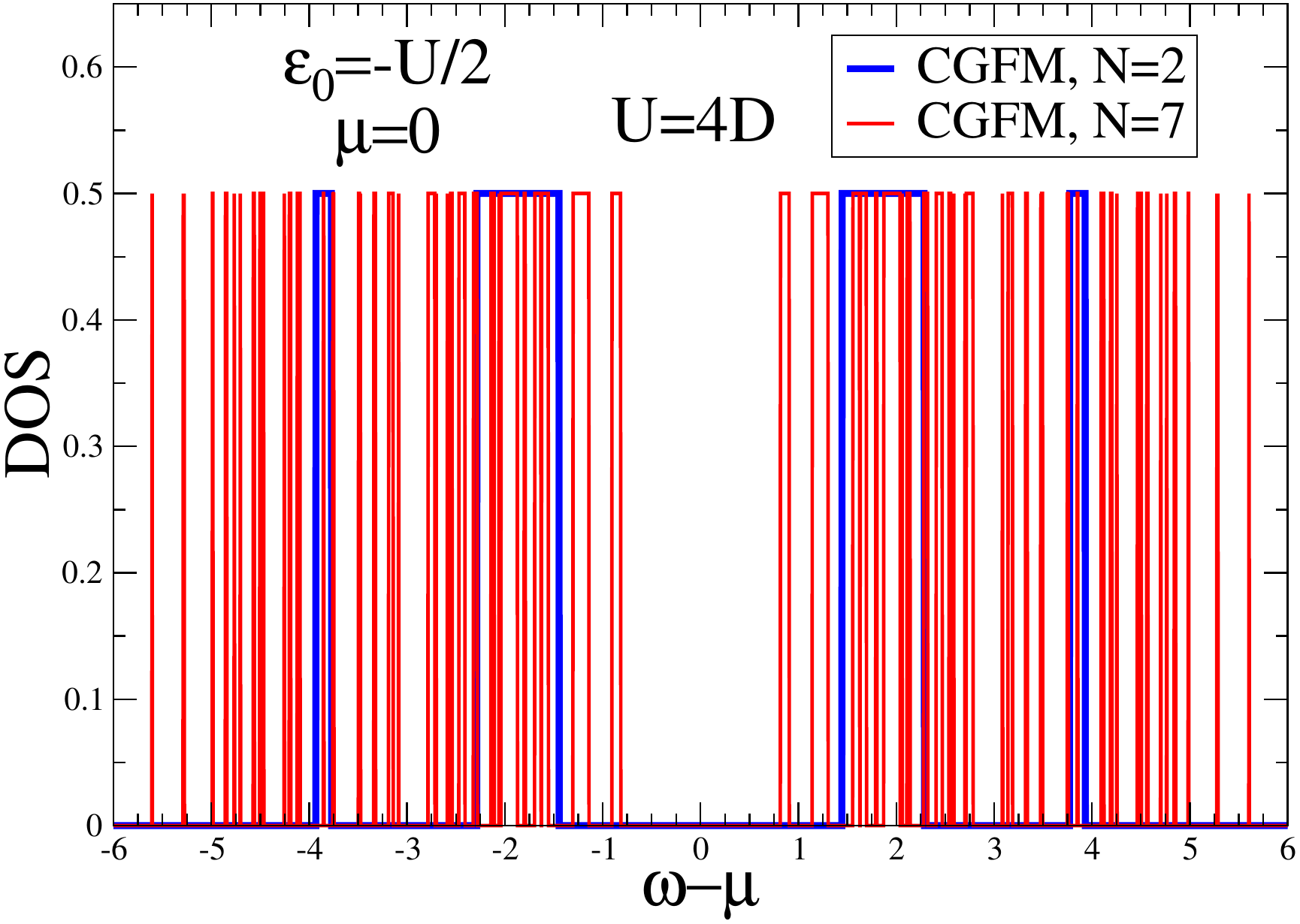}
\caption{Density of states as a function of the frequency $\omega-\mu$ for the half-filling limit, $\epsilon_{0}=-U/2$, $\mu=0$, $T=0.0001t$, $U=4t$ and $N=2$ and $N=7$.}
\label{fig:DOS}
\end{figure}

In the following results we used the temperature $T=0.0001D$, $U=4D$, and magnetic field $h=0$, unless explicitly written otherwise. In Fig. \ref{fig:weight_residues} we represent the residues of the total atomic GF, $g^{at}_{up}=g_{11}+g_{13}+g_{31}+g_{33}$ as a function of the transition energies $u_{i,up}$ for the half-filled limit, $\epsilon_{0}=-U/2$, $\mu=0$, and $N=2,3,4,5,6,7$. The residues exhibit the characteristic mirror symmetry of the one-dimensional Hubbard model \cite{Angela2020}. The striking point here is that as the cluster size increases, the number of atomic residues increases very rapidly (for $N=2$, we have $r_{i,up}=4$; for $N=3$, $r_{i,up}=8$; for $N=4$, $r_{i,up}=32$, and for $N=7$, $r_{i,up}=1322$) while their weight decreases very rapidly. It is worth pointing out that the clusters containing $N=7$ or $N=8$ correlated sites present the best cost-benefit in terms of computational time. They also constitute excellent approximations of the results obtained by the Bethe ansatz for the single-particle gap, the ground-state energy (GSE), the double occupancy, and the phase diagram \cite{Angela2020}.

\begin{figure}[t]
\centering
\includegraphics[clip,width=0.45\textwidth,angle=0.]{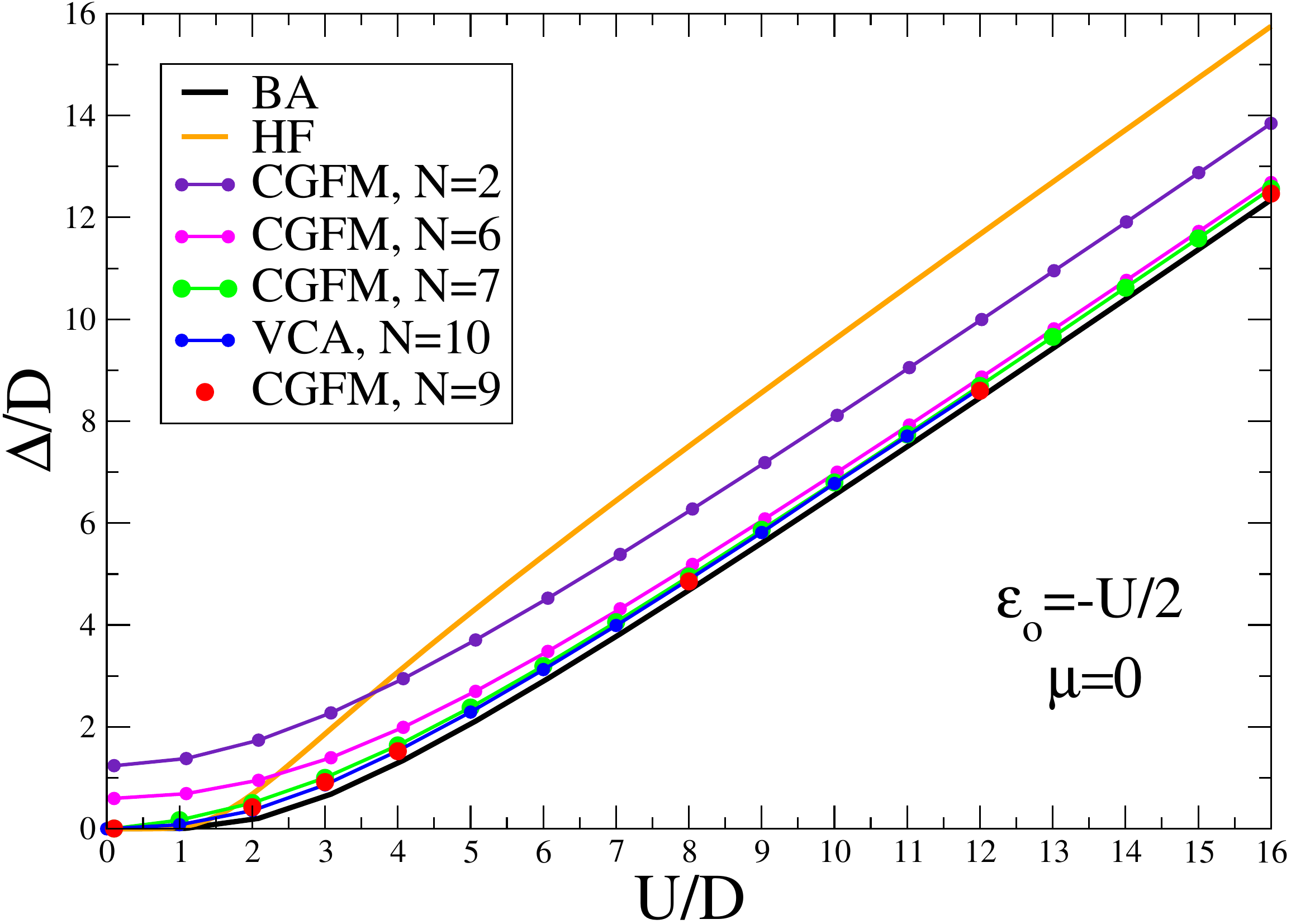}
\caption{Single-particle gap as a function of the correlation energy $ U $ for the BA, the VCA \cite{Potthoff2008} and the CGFM.}
\label{gap_dos}
\end{figure}

\begin{figure}[t]
\centering
\includegraphics[clip,width=0.45\textwidth,angle=0.]{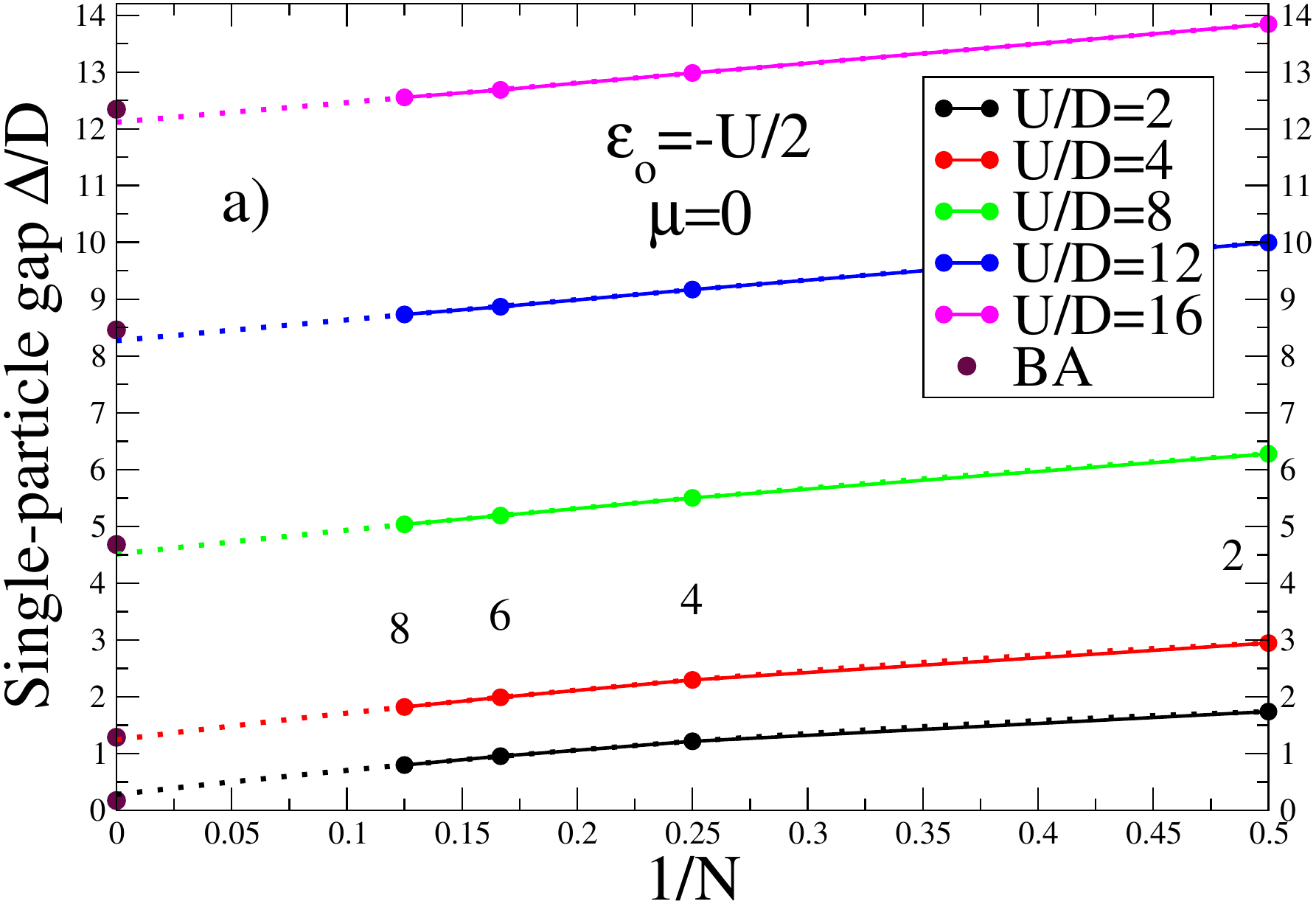}\\
\includegraphics[clip,width=0.45\textwidth,angle=0.]{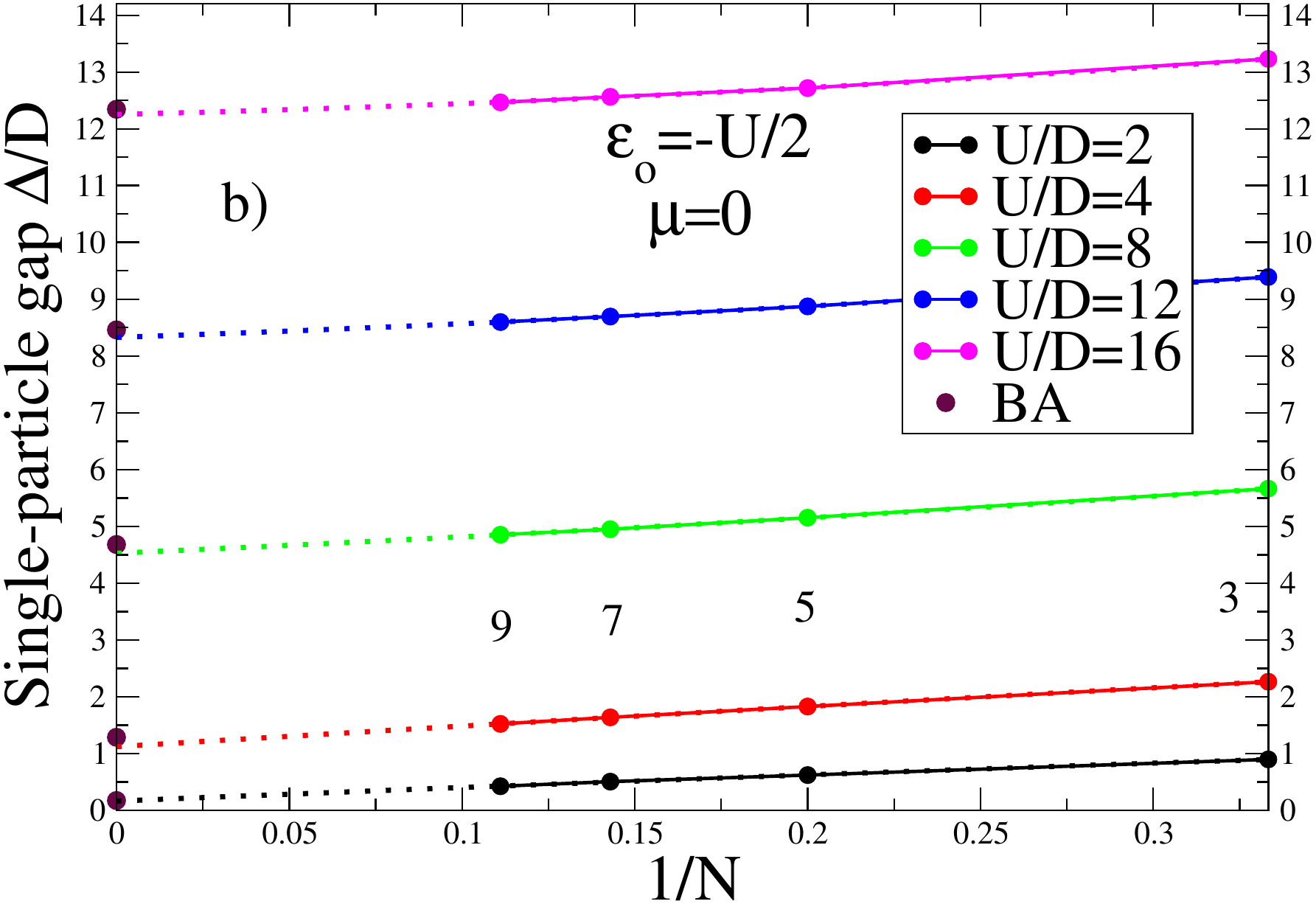}
\caption{Single-particle gap as a function of the inverse cluster size $ 1/N $ for even a) $N=(2,4,6,8)$ and b) odd values: $N=(3,5,7,9)$ for different $U$ values. The pointed lines represent the quadractic regression of the calculated points, and converges to the exact BA results represented  by brown points on the vertical axis. }
\label{Even}
\end{figure}

In figure \ref{fig:DOS}, we plot the DOS as a function of the frequency $\omega-\mu$ for the half-filling limit, $\epsilon_{0}=-U/2$, $\mu=0$, and $N=2$ and $N=7$. The figure presents a discontinuous shape, with regions of different widths separated by gaps, which is a consequence of the atomic cluster employed as a ``seed" to calculate the lattice density of states. As the size of the cluster increases, the DOS tends to become denser and fills all the gaps for $N$ sufficiently large. The gap decreases as the cluster size increases and tends to the BA exact result as indicated in Figs. \ref{gap_dos} and \ref{Even}. It is also worth noting that the particle-hole symmetry is fulfilled for all values of $N$.

\begin{table}
\centering
{\setlength{\extrarowheight}{3.0pt}
{\renewcommand{\arraystretch}{1.0}
 \begin{tabular}{| c | c | c | c |} 
 \hline
 N & $ GSE/N $ & spin ($S_z$) & charge (${\cal Q}$)   \\ [0.5ex] 
 \hline\hline
  2 &  -2.415 & 0  & 2  \\  
 \hline
  3  & -2.413 & $\pm 1/2$ & 3   \\  
 \hline
  4  & -2.488 & 0 & 4  \\  
 \hline
  5  & -2.484 & $\pm 1/2$ & 5 \\ 
 \hline
  6 &  -2.515 & 0 & 6 \\  
 \hline
 7 & -2.511 &  $\pm 1/2$ & 7 \\ 
 \hline
 8 & -2.530& 0  & 8 \\ 
 \hline
 9 & -2.527 & $\pm 1/2$ & 9 \\ 
 \hline
\end{tabular}}}
\caption{Number of correlated sites $N$, $GSE/N$, spin, and charge of the atomic cluster employed in the calculations of the gap at half-filling, Figs. \ref{gap_dos} and \ref{Even}.}
\label{table2}
\end{table}

In Fig. \ref{gap_dos} we present the gap $\Delta$ in the DOS, at half-filling ($\mu=0$), for different cluster sizes $N$, as a function of the electronic correlation $U$. The exact Bethe ansatz results, Eq. \ref{eq:gap_ba}, show that there is no gap for $U = 0$, which is a requirement not satisfied by even $N$ clusters, as indicated in the figure for $N=2,6$. However, this requirement is satisfied for odd clusters sizes, as indicated in the figure for $N=7$. 

For $N$ even, the total spin of the cluster is $S_{z}=0$, and the ground state is nondegenerate; whereas for $N$ odd $S_{z}=\pm 1/2$ and the $GSE/N$ is double-degenerate as indicated in Table \ref{table2}. Even though the even and odd $N$ have such different properties, the $GSE/N$ present an unusual behavior; they are close together in pairs ($2,3$; $4,5$; $6,7$ and $8,9$) and converge in pairs as $N$ increases. Both cluster solutions, when used as ``seeds" to generate the Green's functions of the lattice, are consistent with the Lieb theorem \cite{Tasaki1998}, the GSE of the model are nondegenerate apart from the trivial spin degeneracy, and have total spin 
$S_{tot} = ||S_{z,up}|-|S_{z,down}||$. It should be noted here that the GSE of the cluster does not directly determine the properties of the corresponding infinite system. The cumulants obtained from the clusters of correlated sites are used in the CGFM as bricks to construct the infinite system that can exhibit properties not present in the cluster, like long-range magnetic order or even superconductivity.

The gap results of the CGFM are consistent with other exact diagonalization approaches \cite{Potthoff2008} (see the one referred to as the direct approach) and much better than the Hartree Fock (HF) approximation. As the cluster size increases, the gap gets closer to the BA. We also include in the figure, for the sake of comparison, the result of the variational cluster approach (VCA) \cite{Potthoff2008} for $N=10$ sites. The results obtained by the VCA for a chain containing $N=10$ sites almost agree with the corresponding $N=7$ CGFM and entirely agree with the $N=9$ CGFM. However, a good laptop runs the CGFM for $N=7$ or $N=8$, whereas the VCA needs more computational power due to the self-consistent calculation for the hopping embodied in the method.

In Figs. \ref{Even}(a,b), we plot the single-particle gap as a function of the inverse cluster size, $1/N$, for even and odd $N$ values and $\mu=0$. Since the results present a change in curvature as a function of $1/N$, as indicated in both figures, we performed a quadratic regression to obtain the converged values. The figures indicate that the gap values obtained from even or odd $N$, by increasing the atomic cluster size, converge well to the exact BA result represented in the vertical axis by brown points. The results for the gap for large cluster sizes tend to be insensitive to even or odd clusters size. 

\begin{figure}
\centering
\includegraphics[clip,width=0.45\textwidth,angle=0.]{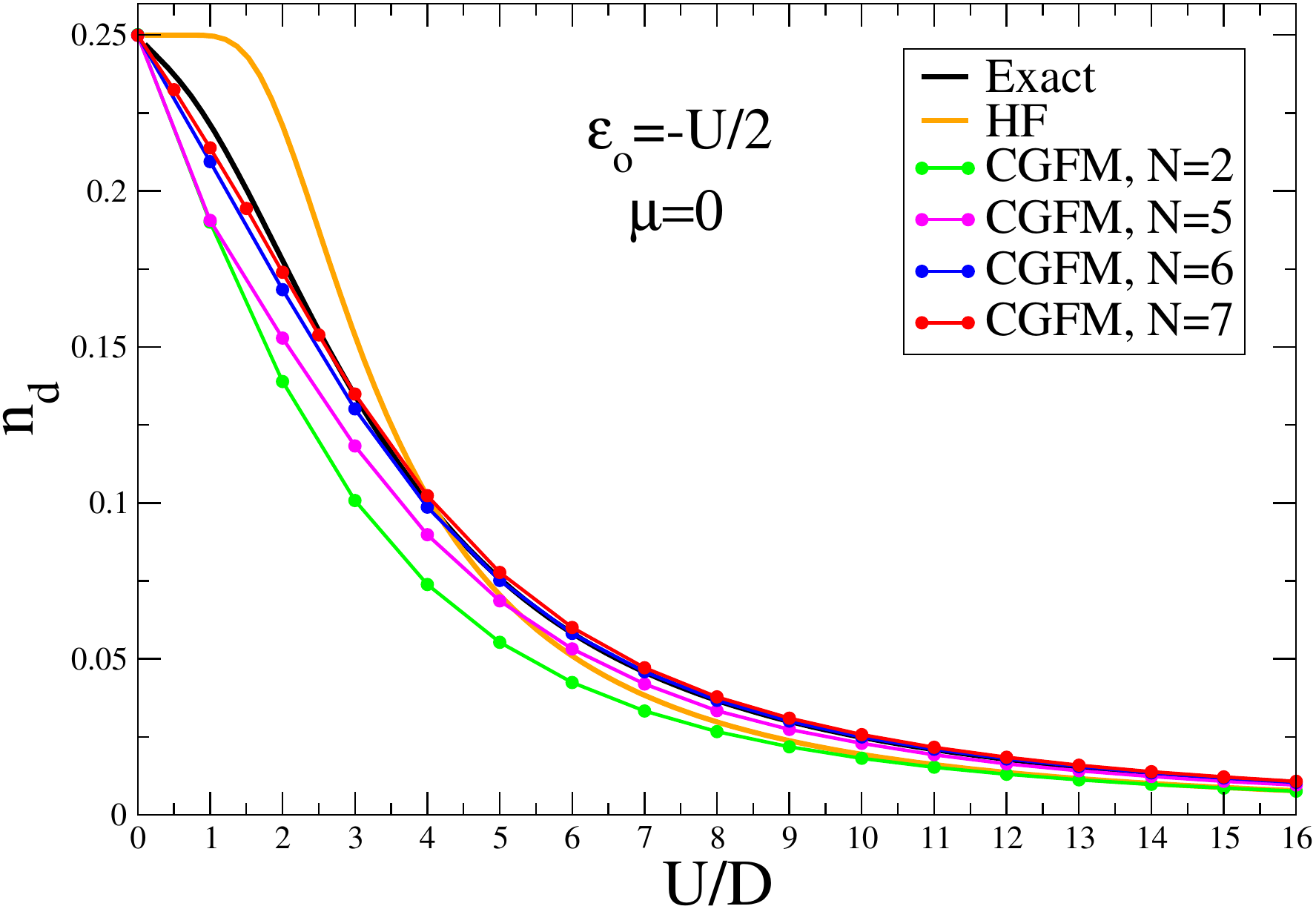}
\caption{Double occupation number $n_d$ as a function of the correlation $U$ for the interpolative ``exact"  result \cite{Essler2005}, the HF approximation and the CGFM for some representative values of $N=2,5,6,7$.}
\label{fig_nd}
\end{figure}
\begin{figure}[t]
\centering
\includegraphics[clip,width=0.45\textwidth,angle=0.]{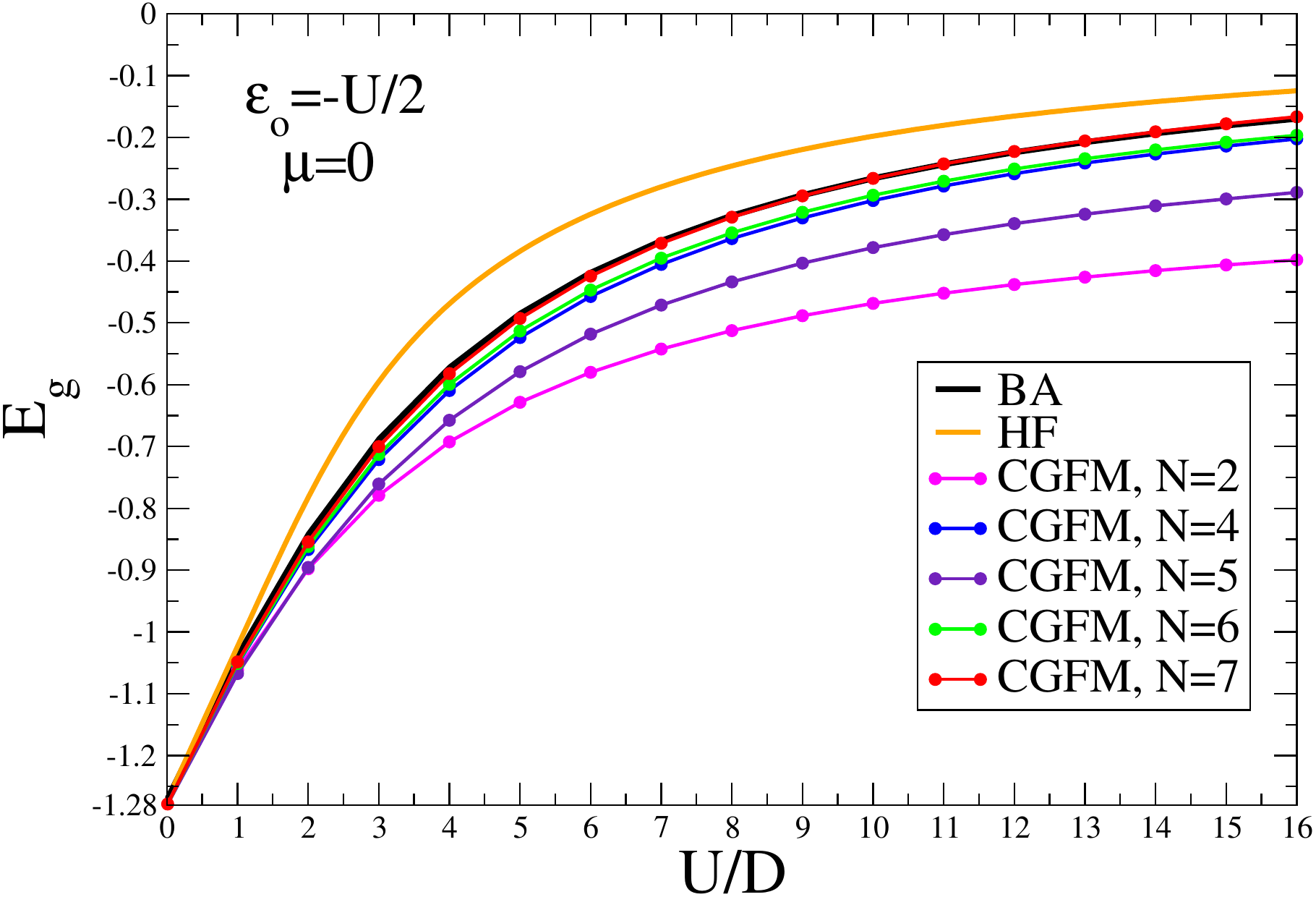}
\caption{Ground-state energy $E_{g}$ as a function of the correlation $ U $ for the BA, the HF approximation, and the CGFM for some representative values of $N=2,4,5,6,7$.}
\label{Ground_state}
\end{figure}

Figure \ref{fig_nd} shows the double occupation number $n_d$ as a function of the correlation $ U $ for an interpolative ``exact" result obtained by the formula, Eq. \ref{Double_occ} \cite{Essler2005}; the HF approximation and the CGFM with $N=2,5,6,7$. The mean-field HF results do not agree well with the interpolative ``exact" result. However, we include it here only to call the attention that the CGFM, contrary to HF, goes to the interpolative result as the number of the correlated sites inside the cluster increases. The CGFM results show that the $N=7$ curve agrees well with the interpolative curve, and only for small $U$ values they present a slight departure. To gather a more precise result in this region, we need to calculate higher-order clusters.

Employing the $n_{d}$ results for different $U$ values and using Eq. \ref{Vignale} we calculate the $E_{g}$ as a function of $U$, as plotted in Fig. \ref{Ground_state}. The exact result is known from the Bethe ansatz formulation, and it was obtained using Eq. \ref{ground-state_ba}. The HF result systematically departs from the exact result and only agrees at small $U$ values. On the other hand, the CGFM for $N=2,4,5,6,7$ approaches the exact result as $N$ increases, and for $N=7$, the agreement with the Bethe ansatz is good. We conclude that the CGFM for $N=7$ or even $N=8$ constitutes a reliable and easy method to treat the one-dimensional Hubbard model. Even if it is possible to improve the results using clusters with $N>7,8$, the computational efforts become increasingly high, and for applications to the one-dimensional Hubbard model, like in quantum dot systems \cite{Carolina2021}, the $N=7,8$ CGFM constitutes a good way to take into account the strong correlations in a systematic way.

\begin{figure}[t]
\includegraphics[clip,width=0.45\textwidth,angle=0.]
{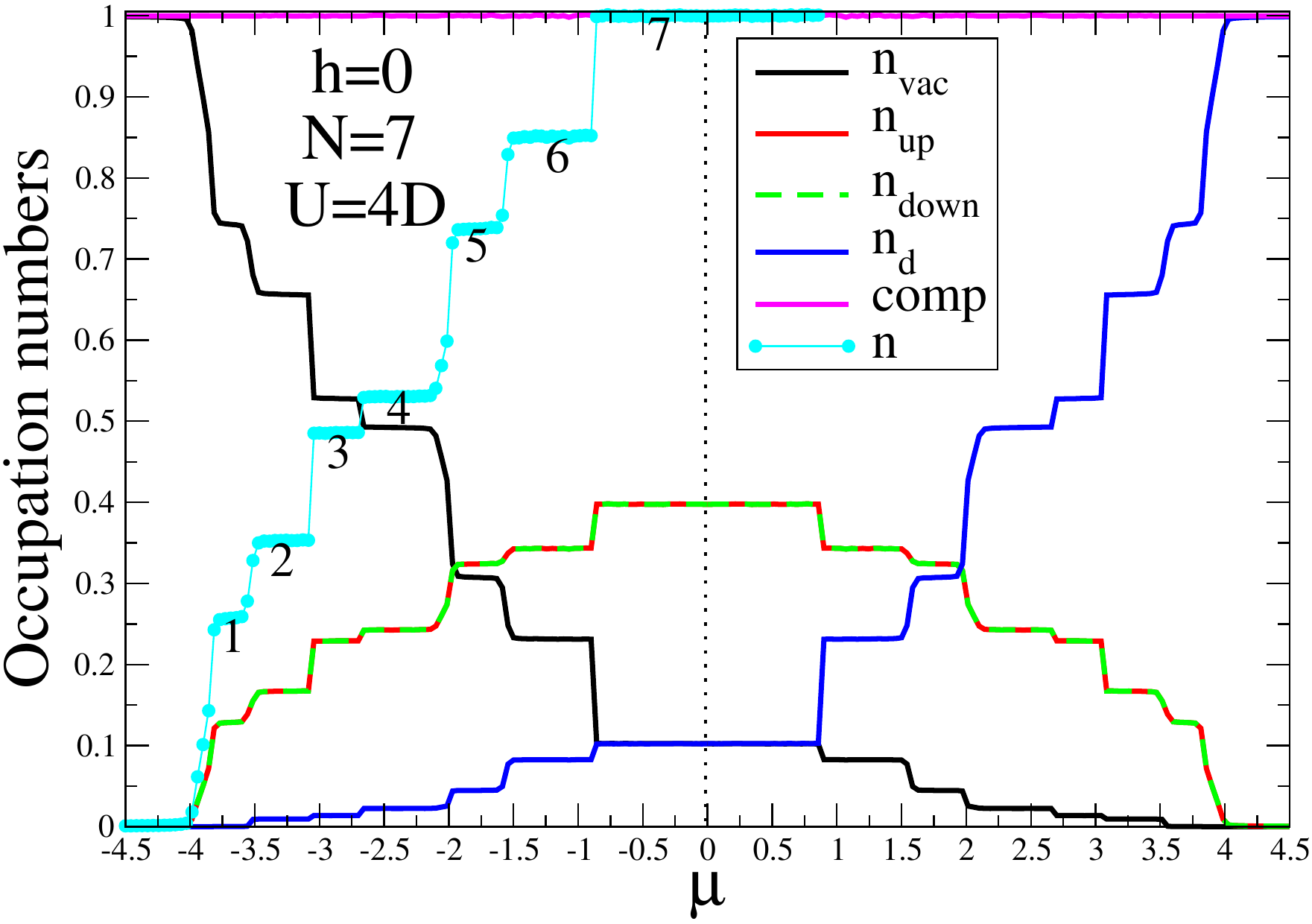}
\caption{Occupation numbers as functions of the chemical potential $\mu$ for the CGFM with $U=4D$, $N=7$ and zero magnetic field.}
\label{Occuph0}
\end{figure}
\begin{table}
\centering
{\setlength{\extrarowheight}{3.0pt}
{\renewcommand{\arraystretch}{1.0}
 \begin{tabular}{| c | c | c | c |} 
 \hline
Region & $\mu$ & spin ($S_z)$ & charge (${\cal Q}$)  \\ [0.5ex] 
 \hline\hline
  1  &  -3.70 &  $\pm 1/2$ &  1  \\ 
 \hline
  2 &  -3.27 & 0  & 2  \\  
 \hline
  3  & -2.90 & $\pm 1/2$ & 3   \\  
 \hline
  4  & -2.40 & 0 & 4  \\  
 \hline
  5  & -1.75 &  $\pm 1/2$ & 5 \\ 
 \hline
  6 &  -1.15 & 0 & 6 \\  
 \hline
 7 & -0.40 & $\pm 1/2$ & 7 \\ 
 \hline
\end{tabular}}}
\caption{Spin and charge of the cluster ground state in different regions when the chemical potential increase from the band's left to the center. }
\label{table1}
\end{table}
In Fig. \ref{Occuph0} we plot the occupation numbers as a function of the chemical potential $\mu$ considering the atomic cluster with $N=7$ correlated sites and the correlation energy $U=4D$. One of the strengths of the CGFM is to discriminate the partial occupation numbers for each spin. We can follow their behavior as the chemical potential varies, changing the occupation density $n$ from small to large values, allowing for a detailed study of the phase transitions present in the model. The figure shows that the completeness relation per spin, Eq. \ref{Occ1}, is satisfied for any value of the chemical potential $\mu$.

Starting from the left of the figure, the vacuum occupation number per spin $n_{\text{vac}}=1.0$, indicating that the system has no electrons at all, and as $\mu$ increases, it goes to zero on the right. The spin up and down occupation numbers $n_{\text{up}}$($n_{\text{down}})$ coincide but they were calculated separately employing the corresponding Green's functions of the sec. \ref{sec3}; both start at zero, go to a constant value at the gap, and then decrease to zero as the system gets entirely doubly populated, $n_{\text{d}}=1.0$.The curves' abrupt character is associated with the change of electron number, which defines the charge ${\cal Q}$ of the cluster, that takes part in the cluster ground state as $\mu$ is varied. In Table \ref{table1}, we show the spin $S_{z}$ and the charge $\cal{Q}$ of the $N=7$ ground state cluster for representative $\mu$ values employed in the calculation of the occupation numbers of Fig. \ref{Occuph0}.  The spin of the ground-state of the cluster alternates between $S_{z}=\pm 1/2$ and $S_{z}=0$,  and the charge starts at ${\cal Q}=1$ at the left, decreases one unit in each step, and arrives at ${\cal Q}=N=7$ at the center of the figure (half-filling). It is also possible to see several flat regions, numbered by $1$ to $7$, where the occupation numbers do not vary, and the system behaves as an insulator. It constitutes an unphysical result, and it is a consequence of the small cluster size employed in the calculation. For large clusters, those insulator regions tend to disappear \cite{Angela2020}. However, there is an exception, the central region, labeled by the number $7$, defines a Mott insulator \cite{Mott49, HubbardIII}.

\begin{figure}[t]
\centering
\includegraphics[clip,width=0.45\textwidth,angle=0.]{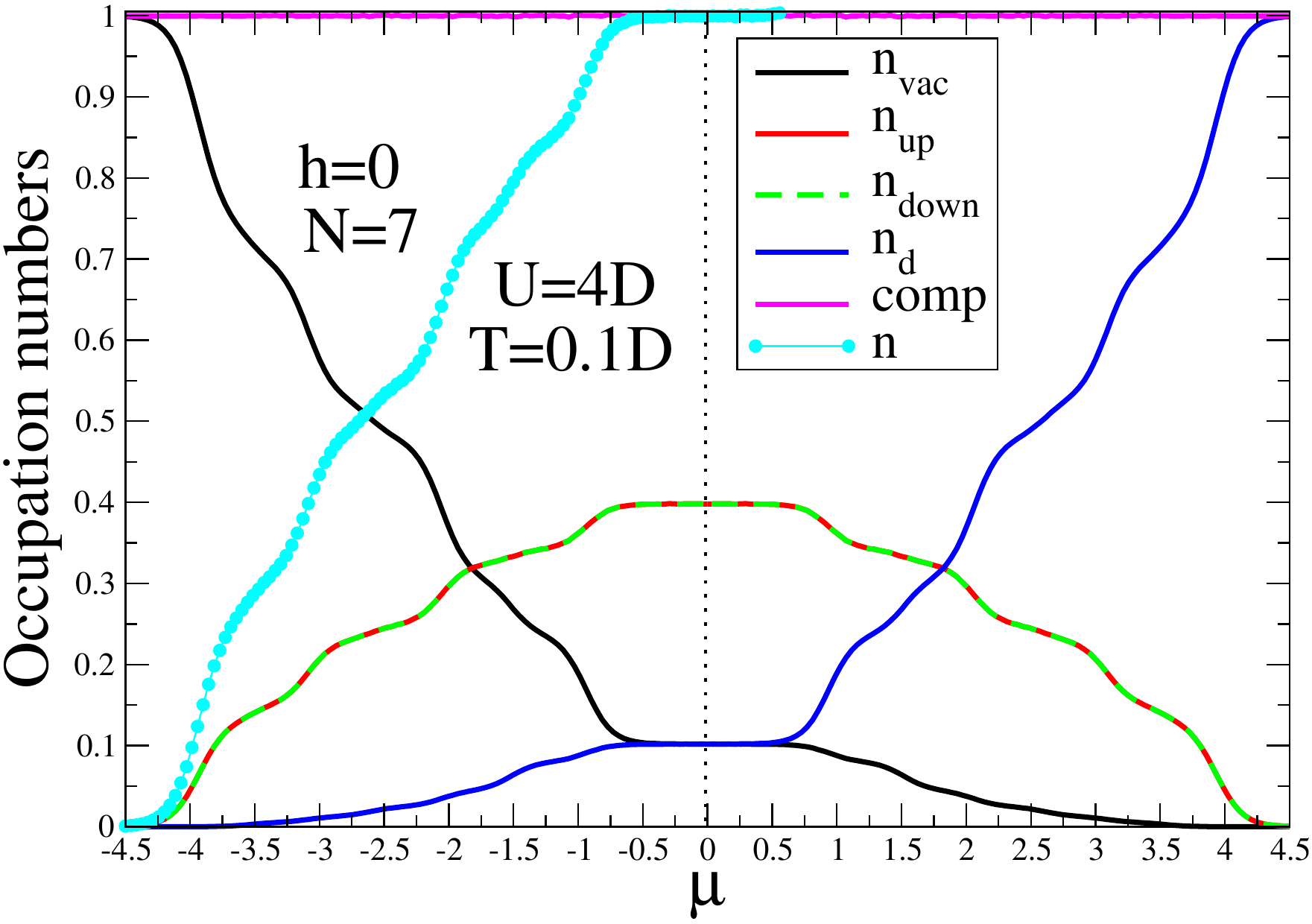}
\caption{Occupation numbers as functions of the chemical potential $\mu$ for the CGFM with $U=4D$, $N=7$, zero magnetic field and $T=0.1D$.}
\label{Occuph0T}
\end{figure}

In Fig. \ref{Occuph0T} we plot the occupation numbers as functions of the chemical potential $\mu$ considering the atomic cluster with $N=7$ correlated sites, the correlation energy $U=4D$, and the temperature $T=0.1D$. The general behavior of the occupation numbers is the same as in Fig. \ref{Occuph0}; however,  the curves are less step-like, and not every transition that satisfies the selection rules will be active for low temperatures. As the temperature increases, the amount of thermal energy in the system activates transitions absent at low temperatures. Their number increases significantly, resulting in smoothing the resulting occupation numbers.

\section{Magnetic field effects}
\label{sec6}

The Hubbard model presents quantum phase transitions at zero temperature \cite{Angela2020}. The ground-state phase diagram in chemical potential vs. magnetic field coordinates was studied in a lengthy way in the reference \cite{Angela2020} employing the quantum transfer matrix method. The model was also studied employing the thermodynamic Bethe ansatz approach \cite{Takahashi1974, Essler2005}, and exhibits five phases \cite{Essler2005, Angela2020} characterized by polarized occupation numbers:

\begin{itemize}

\item{Phase I - Vacuum - $n=0$, $n_{up}=0$, $n_{down}=0$, $m=0$. This phase is characterized by zero occupation numbers. Both bands are empty and the ground-state is the empty lattice. Both electron density and magnetization are zero.}

\item{Phase II - Partially filled and spin polarized band - $0<n<1$, $0 < n_{up} < 1$, $n_{down}=0$, $m=n/2$. The spin-down band is empty and the spin-up band is partially filled.}

\item{Phase III - Half-filled and spin-polarized band - $n=1$, $n_{up}=1$, $n_{down}=0$, $m=1/2$. The spin-down band is empty and the spin-up band is completely filled. The electron density is one and the magnetization is 1/2.}

\item{Phase IV - Partially filled and magnetized band - $0<n<1$, $0 < n_{up} < 1$, $0 < n_{down} < 1$, $0\le m<n/2$. Both bands are partially filled. The electron density is between zero and one and the magnetization is between zero and 1/2.}

\item{Phase V - Half-filled and partially magnetized band - $n=1$, $m\ge 0$. The system is half-filled. The electron density is one, and the magnetization is greater than zero.}

\item{Phase VI - Partially filled and magnetized band - $0<n<1$, $0 < n_{up} < 1$, $0 < n_{down} < 1$, $-n/2\le m<0$. Both bands are partially filled. The electron density is between zero and one, and the magnetization is between -1/2 and zero. This phase is not present on the Bethe ansatz solution, and we believe it is a cluster phase detected at low magnetic fields that should disappear for larger cluster sizes.}

\end{itemize}

\begin{figure}[htb]
\centering
  \begin{tabular}{@{}cccc@{}}
  \includegraphics[width=.23\textwidth]{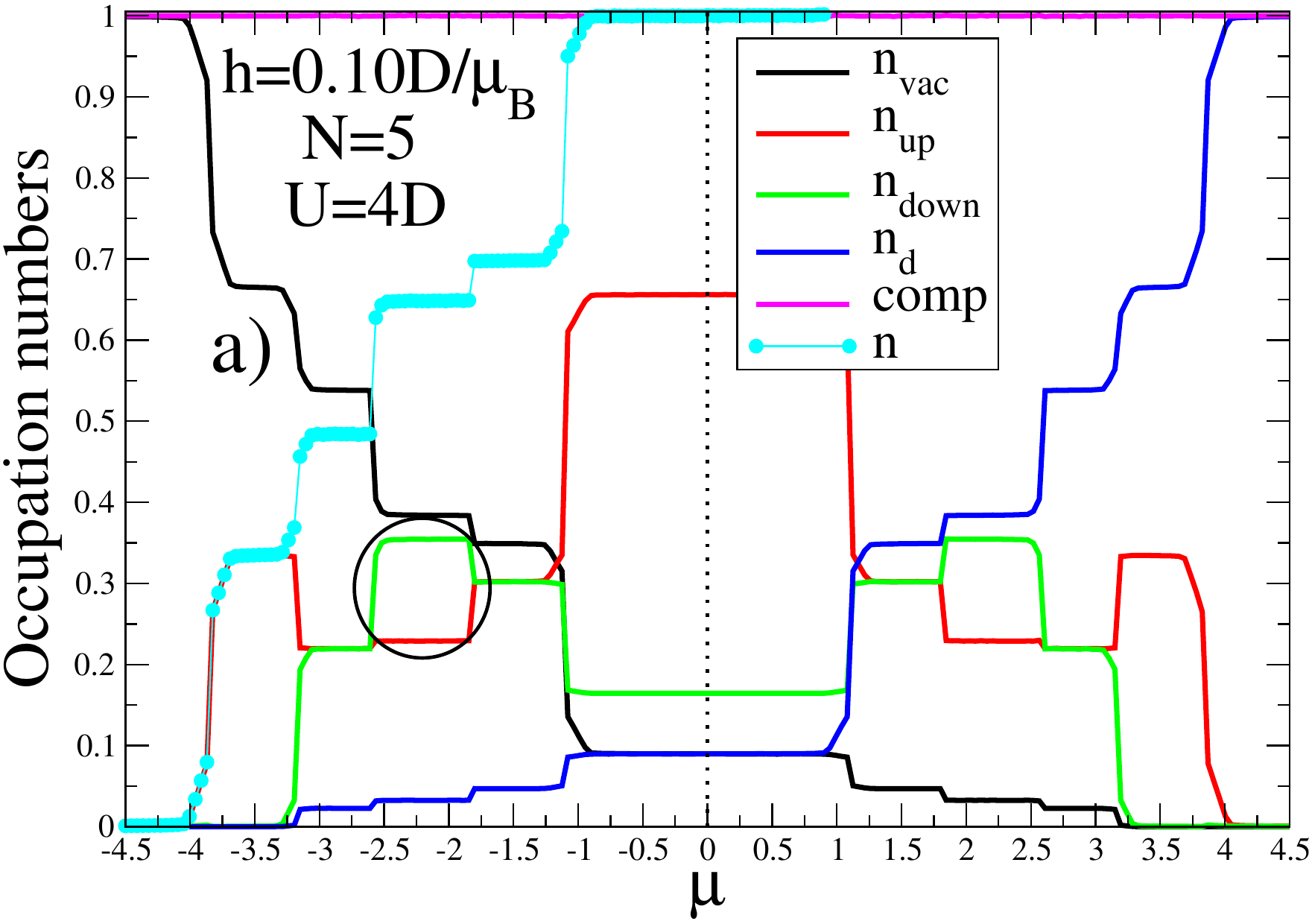} &\
  \includegraphics[width=.23\textwidth]{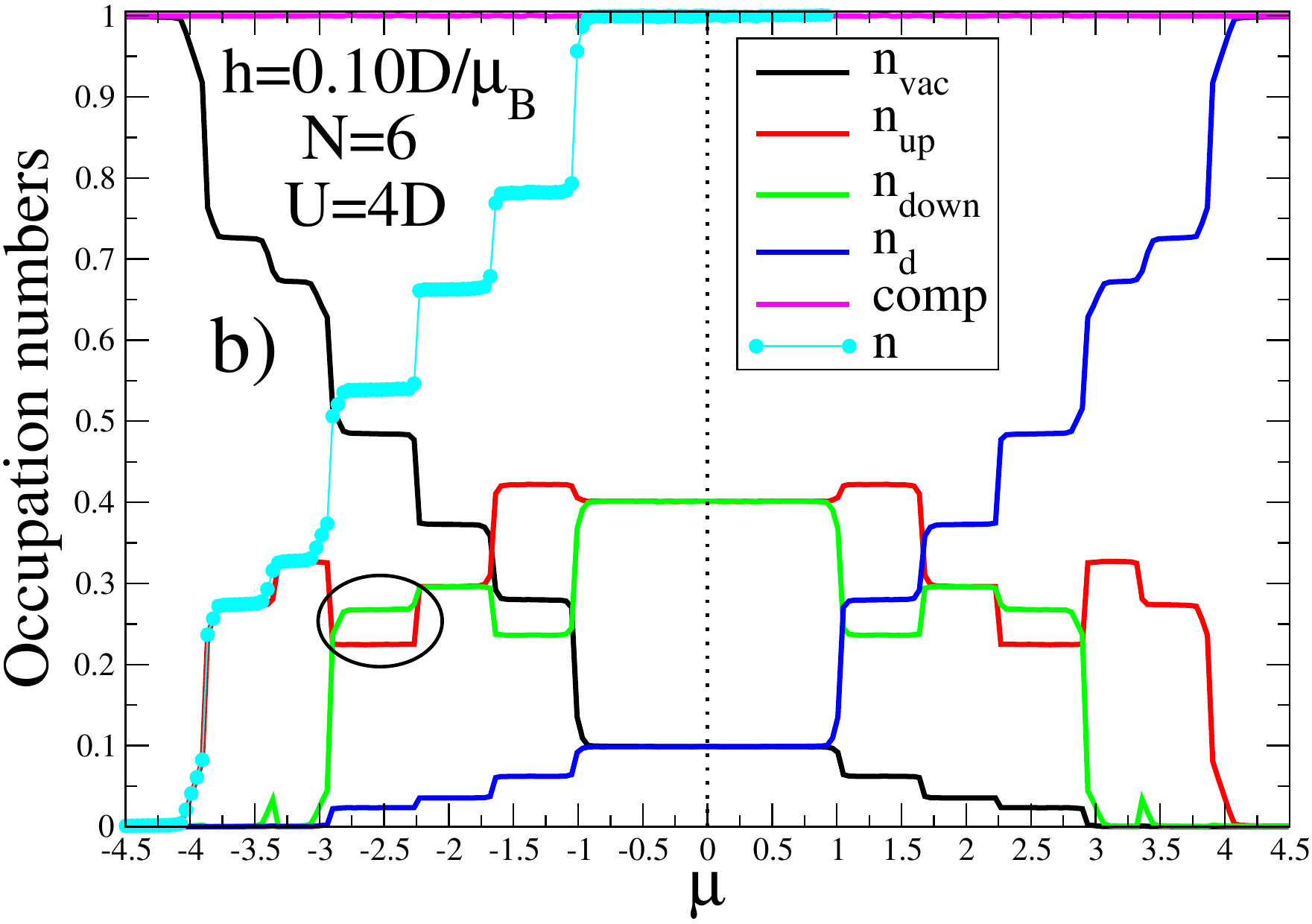}&\\ 
  \includegraphics[width=.23\textwidth]{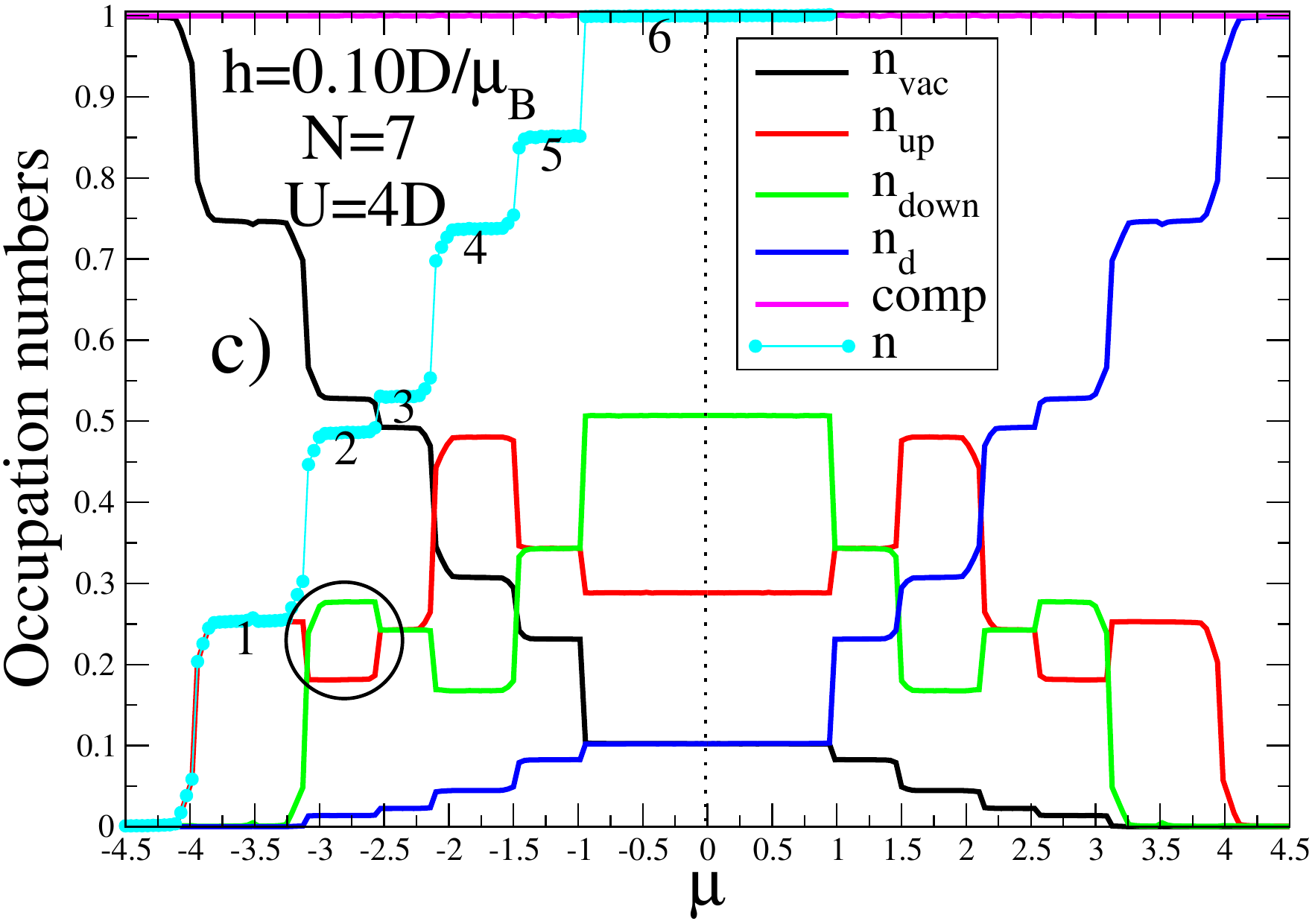} &
  \includegraphics[width=.23\textwidth]{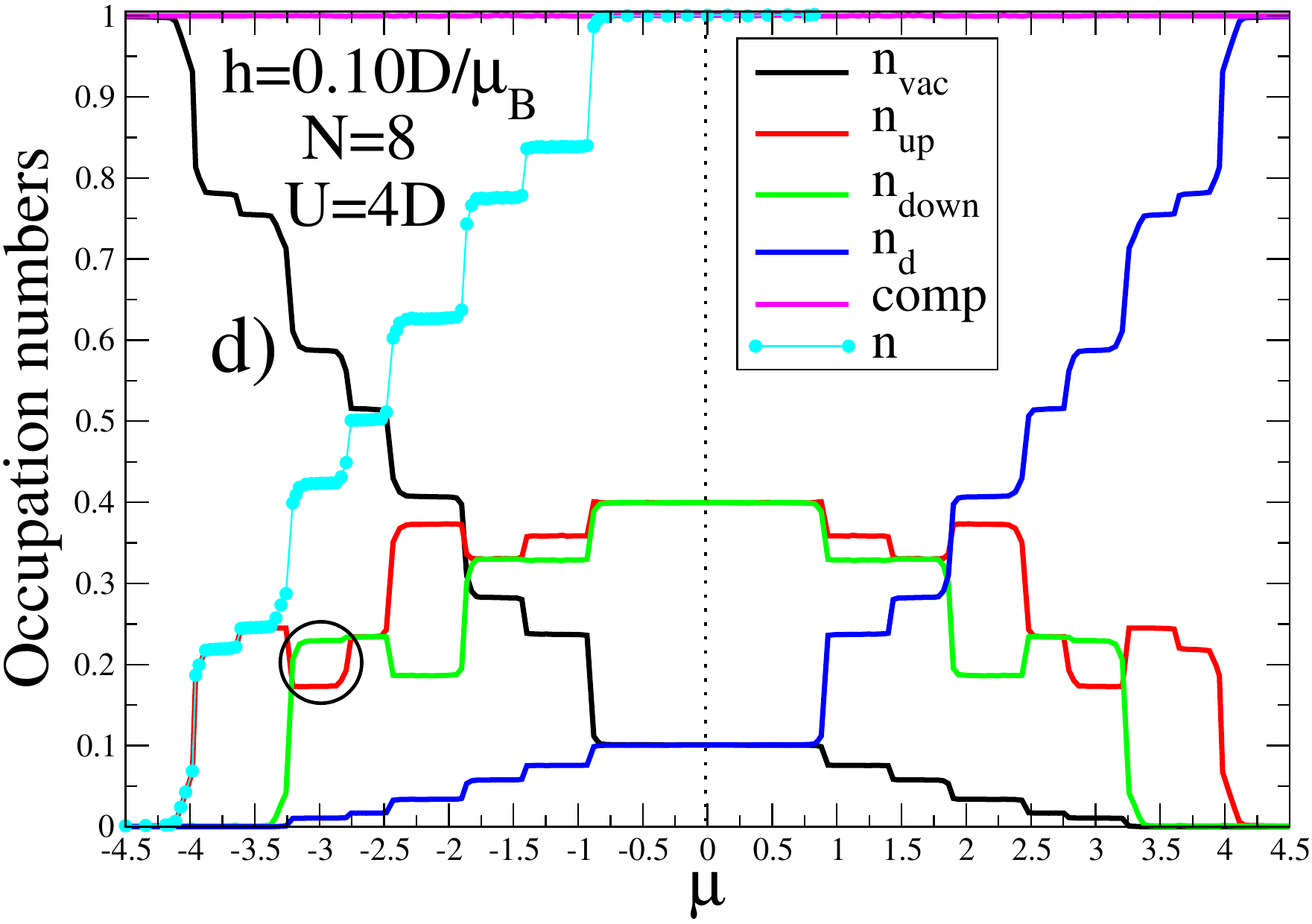}   \\
 \end{tabular}
  \caption{Occupation numbers as functions of the chemical potential $\mu$ for the CGFM with $U=4D$; magnetic field $h=0.10D/\mu_{B}$, and: a)$N = 5$; b)$N = 6$, c)$N=7$, and $N=8$.}
\label{Occuph}
\end{figure}

In Figs. \ref{Occuph}(a,b,c,d) we plot the occupation numbers as a function of the chemical potential $\mu$ for the CGFM with $N = 5,6,7,8$, $U=4D$, and magnetic field $h=0.10D\mu_{B}$. The discontinuities present in the curves result from the small cluster size employed in the calculation, but they tend to disappear with the cluster size increase. The results for $N$ even and odd are completely different, and it is associated with the total spin $S_{z}$ of the cluster ground state: for even $N$, $S_{z}=0$, and $N$ odd, $S_{z}=\pm 1/2$. In the last case, a strong coupling between the spin cluster and the external magnetic field occurs, but this does not happen for even $N$. As $N$ increases, the solutions for both cases tend to the same result because the $S_{z}$ of large odd size clusters tend to be shielded by the other spin's electrons present in the cluster, and the effective $S_{z}$ decreases. The method reproduces all the phase transitions of the model, and the available computational resources establish the limit of the approximate solutions.
\begin{figure}[htb]
\centering
  \begin{tabular}{@{}cccc@{}}
  \includegraphics[width=.23\textwidth]{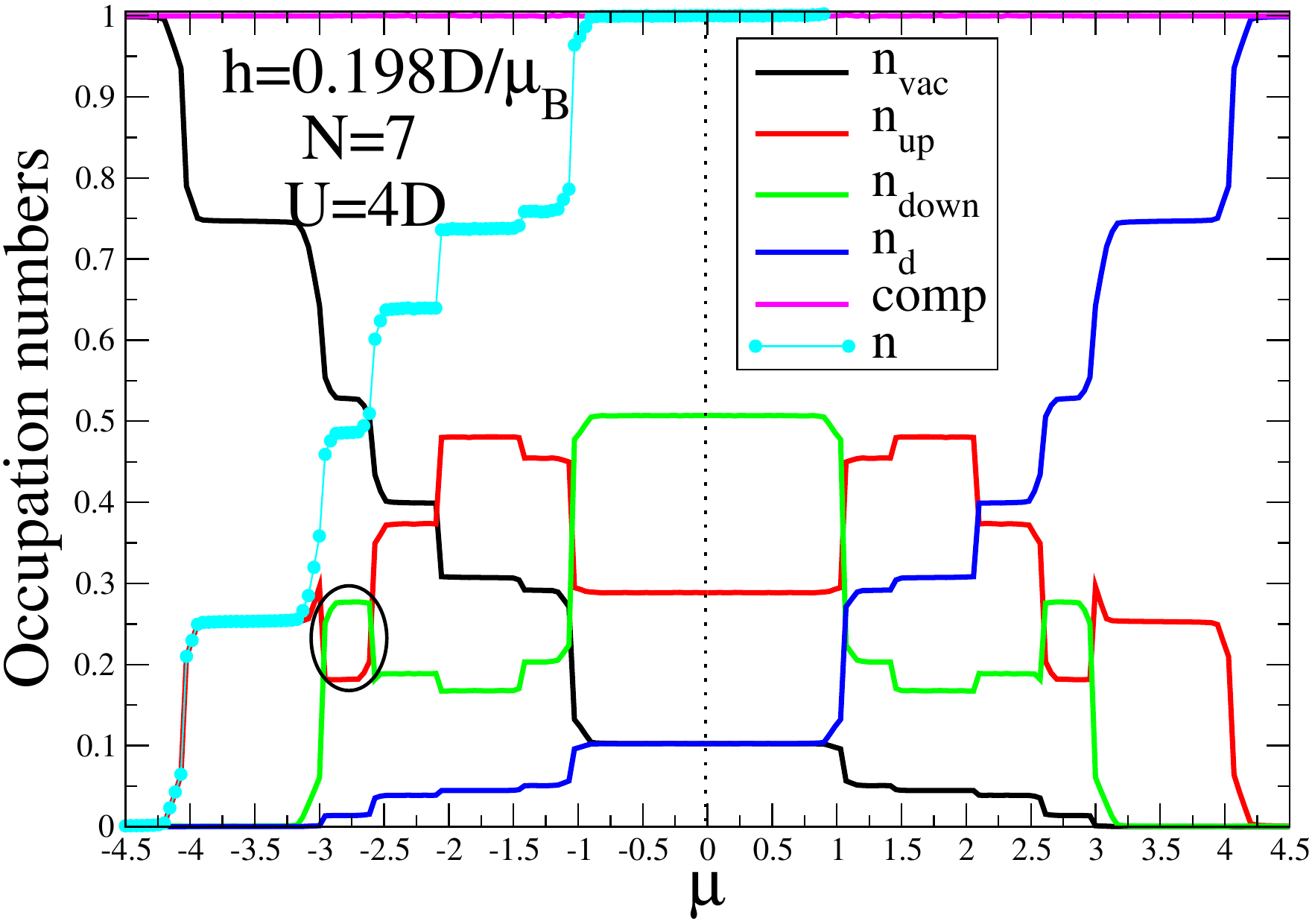} &
  \includegraphics[width=.23\textwidth]{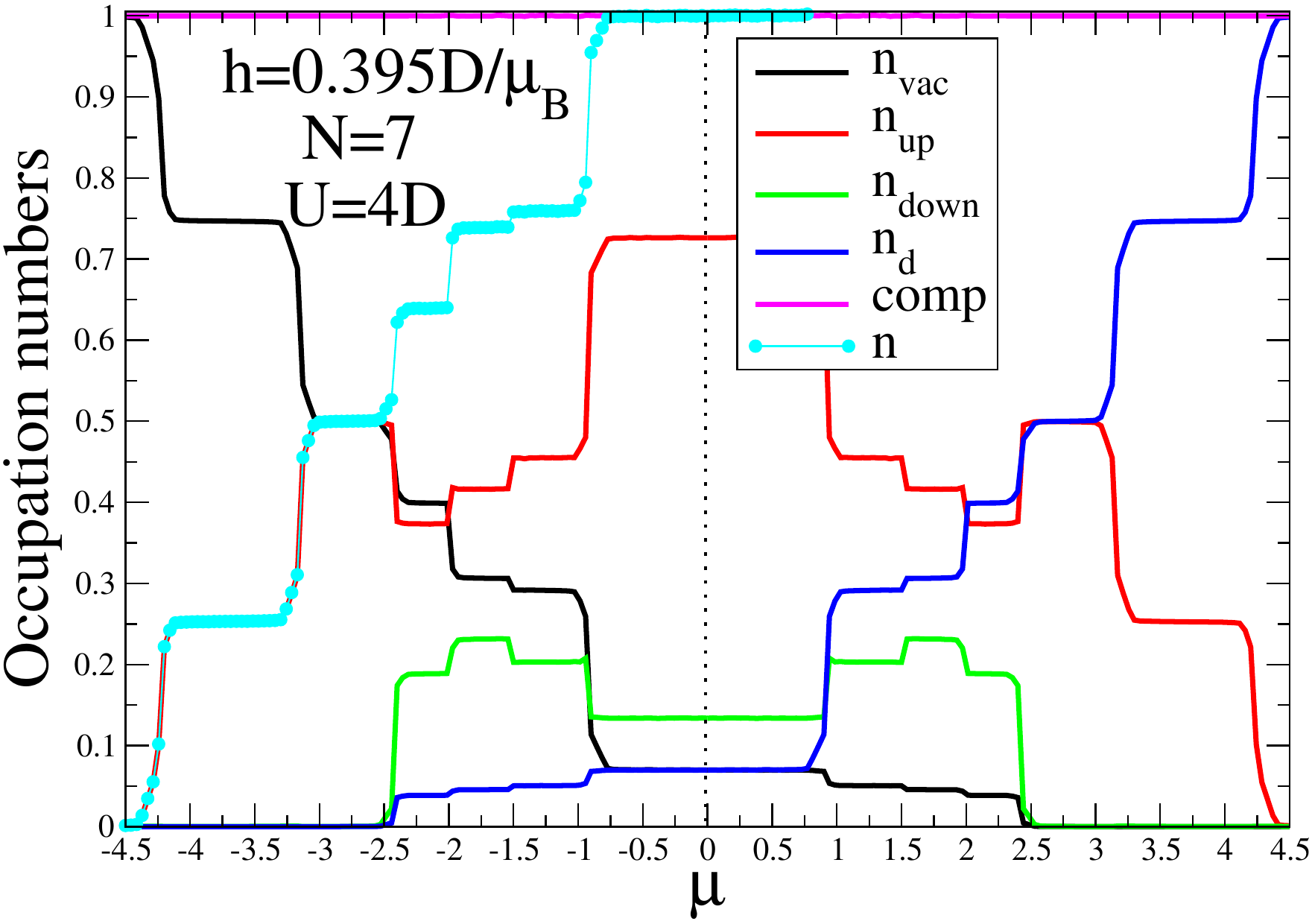} \\
  \includegraphics[width=.23\textwidth]{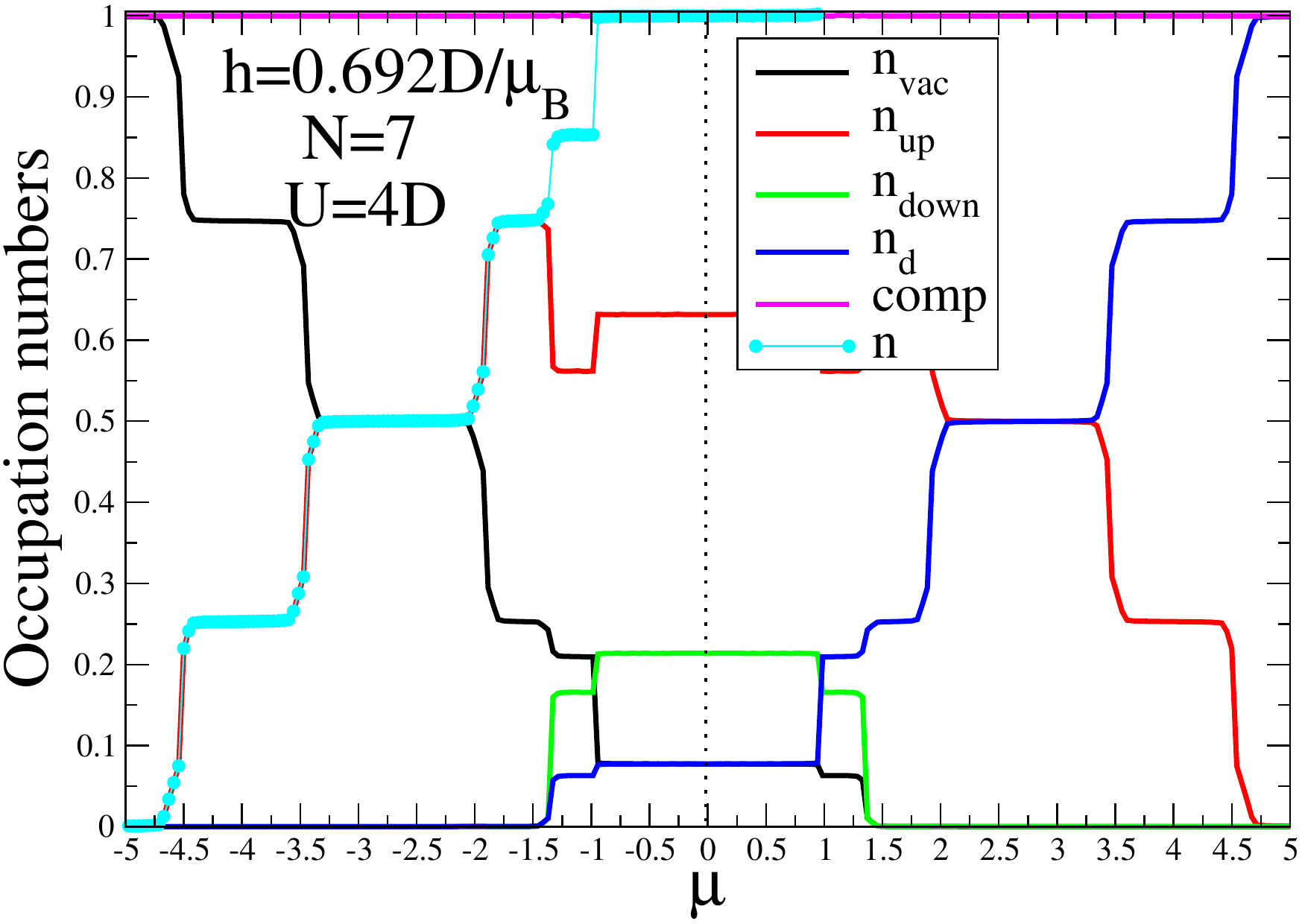} &
  \includegraphics[width=.23\textwidth]{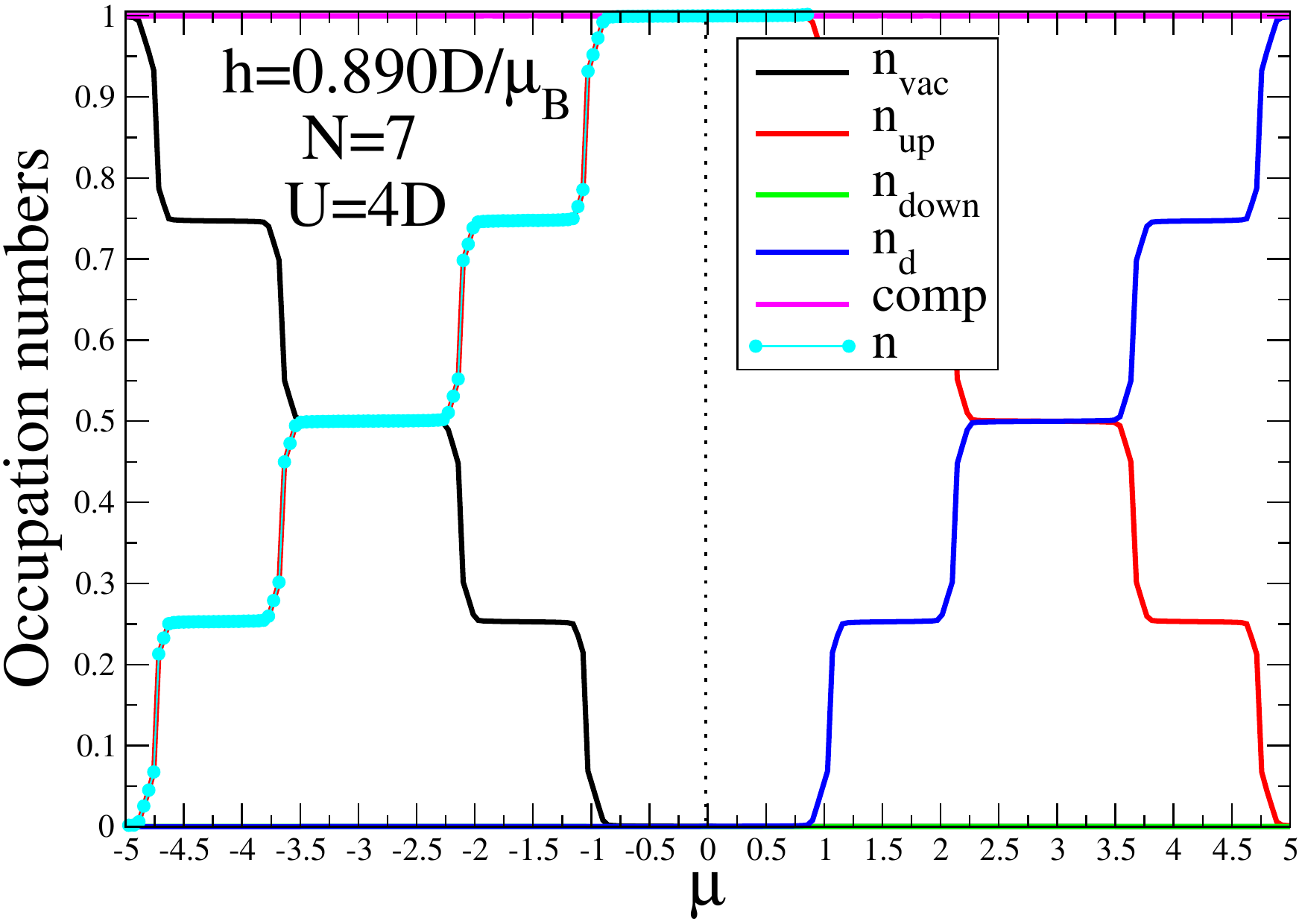}   \\
 \end{tabular}
  \caption{Occupation numbers as functions of the chemical potential $\mu$ for the CGFM with $U=4D$; $N=7$ and magnetic fields: a) $h=0.198D/\mu_{B}$;  b) $h=0.395D/\mu_{B}$;
c) $h=0.692D/\mu_{B}$, and d) $h=0.890D/\mu_{B}$.}
\label{Occuphighh}
\end{figure}
\begin{figure}[htb]
\centering
  \begin{tabular}{@{}cccc@{}}
 \includegraphics[width=.45\textwidth]{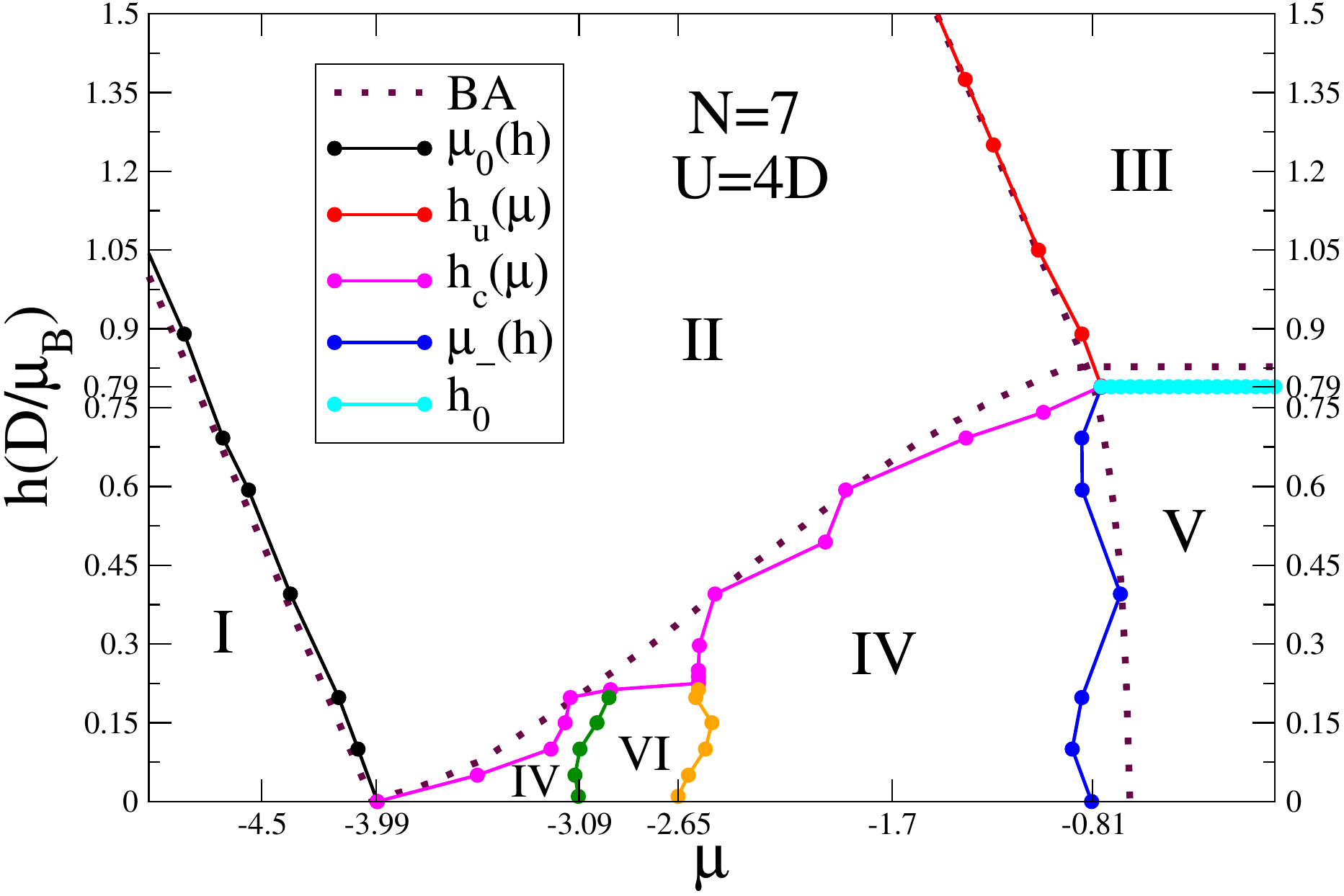} \\
 \includegraphics[width=.45\textwidth]{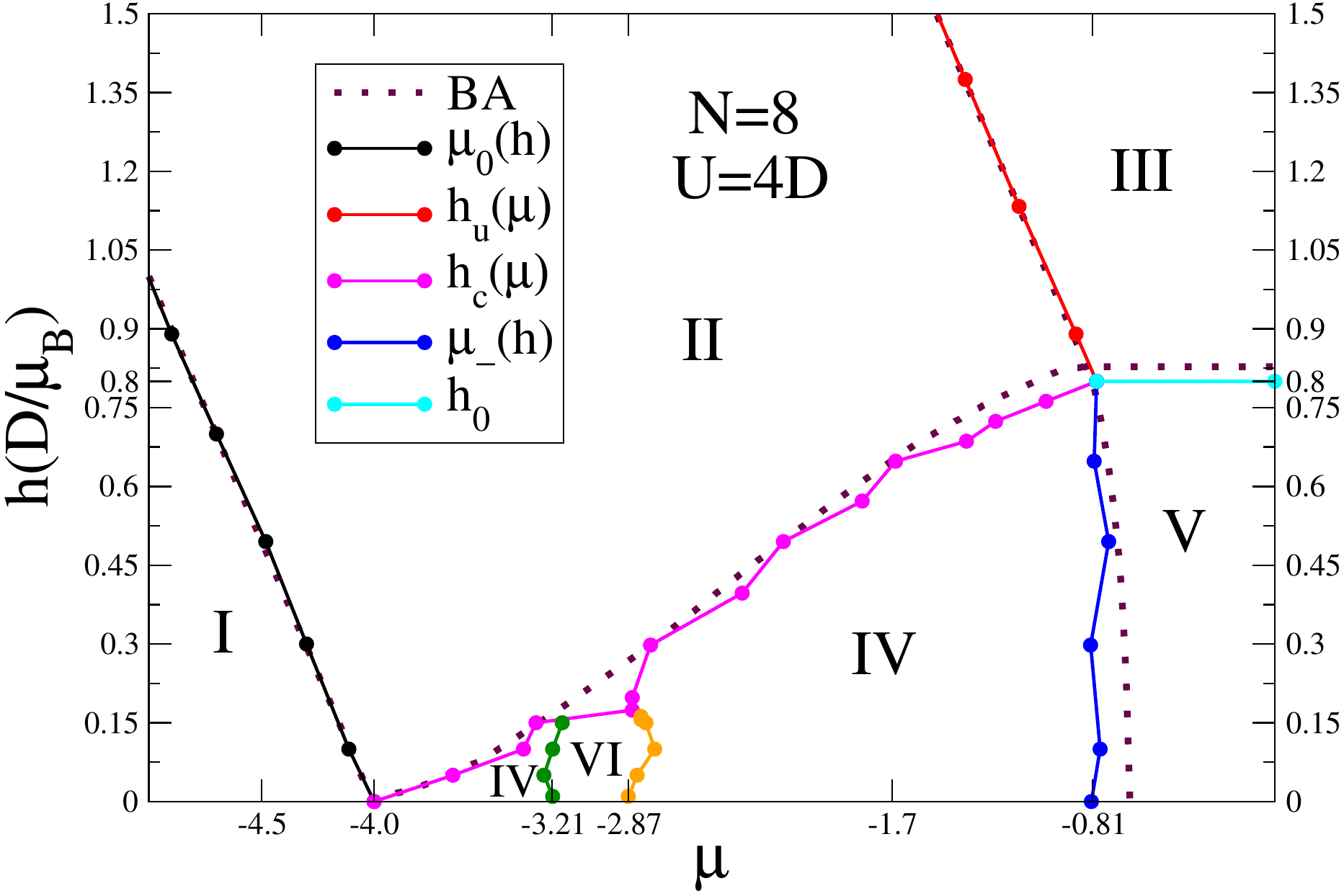} 
 \end{tabular}
  \caption{Ground-state phase diagram in $\mu$ vs. $h$ coordinates for
$U=4D$, $N=7$ and $N=8$. The regions I to VI correspond to different phases of the model. The dotted curves correspond to the exact result \cite{Angela2020}.}
\label{Phase_diagram}
\end{figure}

In Fig. \ref{Occuph}c), due to the action of the magnetic field $h$, there are regions where the CGS total spin is $S_{z}=\pm 1/2$ (regions $1,2,4,6$),  and regions where the CGS has total spin $S_{z}=0$, (regions $3,5$). In regions where $S_{z}=0$, the $n_{up}$ and $n_{down}$ occupations are almost equal, but when $S_{z}=\pm 1/2$, in some regions $n_{up} > n_{down}$ and in others $n_{up} < n_{down}$. In regions $2,6$, $n_{down}>n_{up}$, what is an unexpected result once the magnetic field $h$ tends to favor the $n_{up}$ solution. In region $6$, for $N=5$, $n_{up} > n_{down}$, for $N=7$, occurs an inversion $n_{up} < n_{down}$, and for $N=9$, we have calculated only one point in this region (not showing here) and occurs another inversion for $n_{up} > n_{down}$. Besides this, $n_{up}$ and $n_{down}$ become closer as $N$ increases for the odd cases. The low magnetic fields act in a non-uniform way as a function of the chemical potential, which leads to the generation of this new behavior in the occupation numbers. We detect the same behavior in all curves for $N=5,6,7,8$. We associate to this behavior a ``cluster phase" that we call phase VI, in which the band is partially filled and magnetized - ($0 < n < 1$ and $m < 0$). However, this cluster phase should dissapear for large clusters $N$ size.

Fig. \ref{Occuphighh} show the occupation numbers as functions of the chemical potential $\mu$ for the CGFM with $U=4D$; $N=7$ and magnetic fields: a) $h=0.198D/\mu_{B}$; b) $h=0.395D/\mu_{B}$; c) $h=0.692D/\mu_{B}$, and d) $h=0.890D/\mu_{B}$. The cluster phase VI remains for $h=0.198D/\mu_{B}$, in two regions: the small circled region and  other large region at the Mott gap in Fig. \ref{Occuphighh}a). For intermediate to high magnetic fields the $n_{up}$ is higher than $n_{down}$ independent of the $\mu$ position. On the other hand, the $n_{down}$ active region decreases in direction to $\mu=0$, until disapear at a critical line defined by $h=0.79D/\mu_{B}$ as indicated in Fig. \ref{Phase_diagram}, where the system suffers a transition to phases II or III depending on the $\mu$ position.

Fig. \ref{Phase_diagram} shows the ground state phase diagram in $\mu$ vs. $h$ coordinates for $U=4D$, $N=7$ and $N=8$. The regions I to VI are characterized by different values of the density $n$, partial occupation numbers $n_{up}$ and $n_{down}$ and the magnetization $m$. The exact boundary result of phases I and II is $\mu_{0}(h)=-2-2u-h$ \cite{Angela2020} and the CGFM results converge to the $h=0$ limit ($\mu_{0}=-4.0$) as we increase the number of sites, and for $N=8$, $\mu_{0}=-4.0$. In the same way, the exact boundary result of phases V and III is $h_{0} =2(\sqrt{1+u^{2}})-2u=0.828$, and the corresponding results of CGFM is for $N=8$, $h_{0}=0.8$. The exact boundary transition from IV to V region is given by $\mu_{-}(h=0)=0.64$ \cite{Angela2020} , and the results of CGFM for $N=8$, is 
$\mu_{- }(h=0)=-0.81$.  The results of the boundary transitions converge to the exact one as we increase $N$, and the phase diagram exhibits the same shape, the same phases, and agrees well with the one obtained from exact methods such as the TBA and QTM methods \cite{Takahashi2002, Campo2015, Angela2020, Sacramento2021, Juttner1998}.

\section{A simple application}
\label{sec7}

\begin{figure}[t]
\centering
\includegraphics[clip,width=0.45\textwidth,angle=0.]{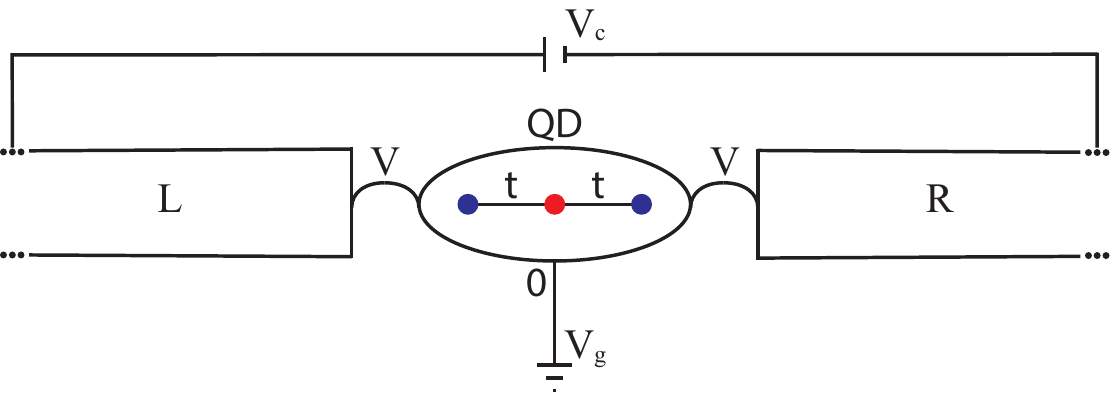}
\caption{Schematic view of a correlated 3-site quantum dot immersed in a left (L) and right (R) correlated 3-site Hubbard rectangular conduction bands.}
\label{SET}
\end{figure}

This section shows a simple application of the CGFM in spintronics. We study the electronic transport through a quantum wire (QW) described by a correlated 3-site Hubbard rectangular conduction leads with an immersed correlated 3-site quantum dot. Fig. \ref{SET} shows a schematic view of of the setup \cite{Carolina2021}. When connected to Hubbard leads, the CGFM clusters can be used as correlated quantum dots to realize a single-electron transistor (SET). The cluster can be viewed as a complex level structure that works as a QD. Using the gate voltage $V_{g}$ we can tune the alignment of the different energy levels of the QD with the chemical potential $\mu$ to realize the polarization of the spin current that is established through the device by the voltage $V_{c}$. The coupling of the Hubbard leads to the QD is given by a hopping term $V$ that transfers electrons from the leads in and out of the dot.

The local Green's function of the dot connected to the two 3-site Hubbard rectangular conduction leads is given by~\cite{Odashima2017},
\begin{equation}
\label{condut}
G_{\sigma}^{00}(\omega,T)=\frac{g_{QD}(\omega,T)}{1-2|V|^2 g_{QD}(\omega,T) G_{\sigma}(\omega,T)},
\end{equation}
where $g_{QD}(\omega,T)$ is given by Eq. \ref{eq:771} and the leads Green's functions, $G_{\sigma}(\omega,T)$, by Eqs. \ref{eq:992} and \ref{eq:222}. The dimensionless conductance of the device can be calculated employing the standard relation \cite{Dong_02}
\begin{equation}
\label{sigma}
G/G_{0}=\int d\omega \left(-\frac{\partial f}{\partial \omega}\right)\mathcal{T}(\omega,T) ,
\end{equation}
where $G_{0} = 2e^{2}/h$ is the quantum of conductance (taking spin into account),   $f(\omega)$ is the Fermi-Dirac distribution and
$\mathcal{T}(\omega,T) =\Gamma Im(G^{00}_{\sigma}(\omega,T))$ is the transmittance,
with $\Gamma= 2 |V|^2 Im(G_{\sigma}(\omega,T))$.
\begin{figure}[t]
\centering
\includegraphics[clip,width=0.45\textwidth,angle=0.]{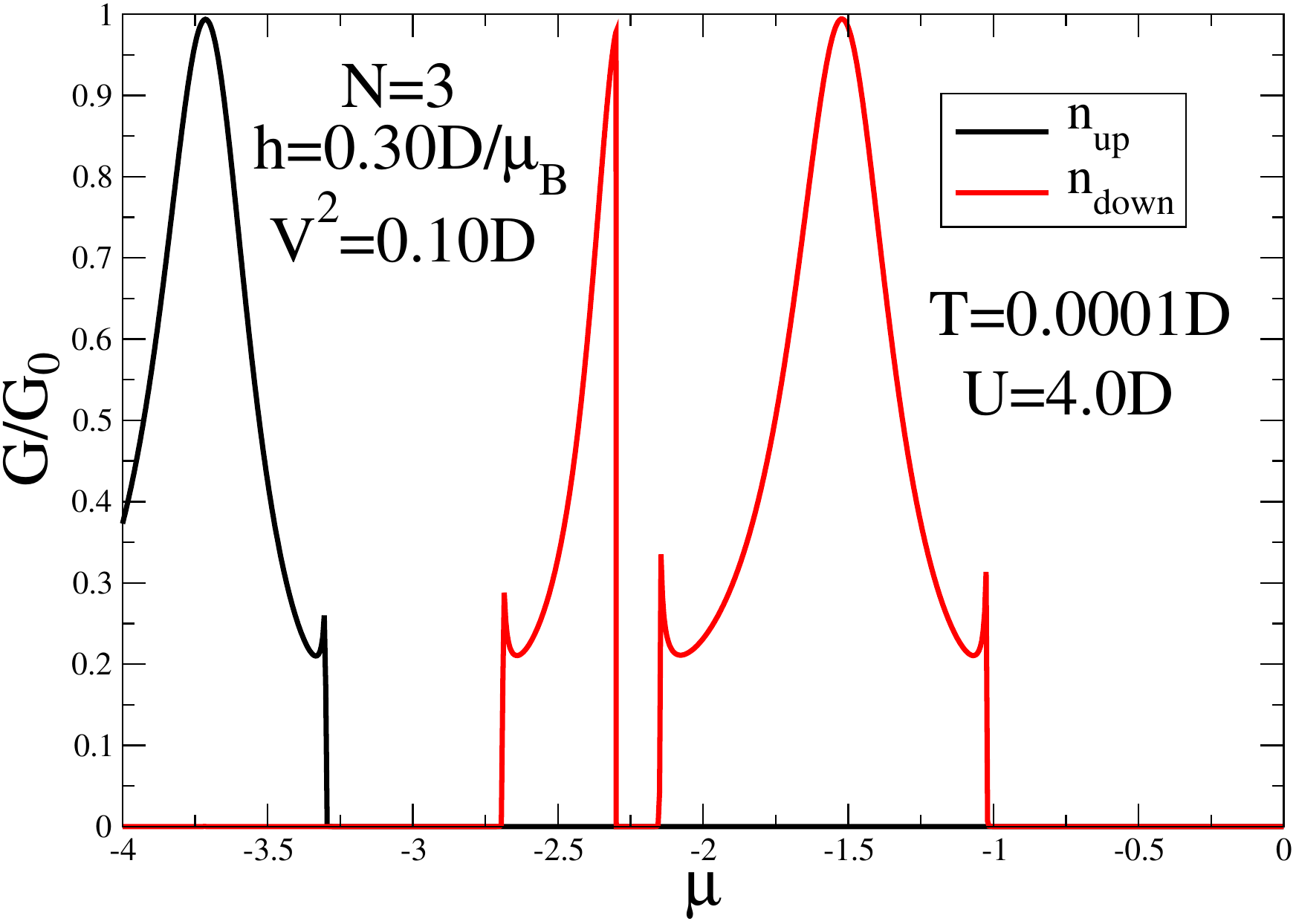}
\caption{Electrical conductance $G/G_{0}$ vs. chemical potential $\mu$.} 
\label{Cond}
\end{figure}

Fig. \ref{Cond} shows the polarized $G/G_{0}$ vs. $\mu$. For simplicity, we consider the trimer, $N=3$, and the parameters employed in the calculation are: $T=0.0001D$, $U=4.0D$, $V^{2}=0.10D$, and the magnetic field $h=0.30D/\mu_{B}$. There is a region, approximately in the interval $[0.29,0.42]D/\mu_{B}$, where we have well-defined polarized spin currents; for other values of $h$, there are components of spin up and down currents corresponding to the same $\mu$. Since these QDs interact strongly with an external magnetic field generating polarized spin regions as a function of $\mu$, they can be valuable for applications in spintronics.

\section{Conclusions and Perspectives}
\label{sec8}

We developed a method to solve the single-band Hubbard Hamiltonian employing cumulants to construct the Green's functions of the lattice, the CGFM. The method focuses on a cluster solution (``seed"), employing ED techniques, and can be extended to different strongly correlated systems: Anderson Hamiltonian, the $t-J$, Kondo, and Coqblin-Schrieffer model. The method is sufficiently general to be applied to any model parameter space and in 1D, 2D, and 3D systems. One central point of the CGFM that differentiates it from other exact diagonalization approaches like the VCA \cite{Potthoff2008, Enrique2008, Seki2018}, is the calculation of all atomic Green's functions employing the Lehmann representation. It constitutes the hard part of the method. It allows the investigation of the relevant physical processes in each parameter space of the Hubbard model, providing clues to clarify the different ground states present in the model.

The calculations embodied inside the method are direct, and no self-consistency is needed. We present the mathematical derivation of the formalism and apply it to the single-band one-dimensional Hubbard Hamiltonian. We benchmarked the results of the CGFM against the exact one obtained with the thermodynamic Bethe ansatz and the quantum transfer matrix method \cite{Takahashi2002, Campo2015, Angela2020, Sacramento2021, Juttner1998}. We calculated the gap in the density of states, the ground state energy, and the occupation number density. Systematically, all the results tend to be exact as the number $N$ of correlated sites inside the cluster increases. The precision of the approximations depends on the computational resources available. It is possible to use the CGFM by employing clusters of $N=7$ or $N=8$ correlated sites in a robust personal computer, whereas more sites require heavy computation systems.

We recovered all the five phases  exhibited by the one-dimensional Hubbard model in the presence of a magnetic field. We calculated the ground-state phase diagram in $\mu-h$ coordinates for $N=7$ and $N=8$ for $U=4D$. The regions I to V are characterized by different density values $n$ and agree well with the exact phase diagram. In addition to this, we identified a cluster phase (phase VI) that exists only for low magnetic fields and corresponds to a partially filled band $0 < n < 1$, but with $n_{down}>n_{up}$ and a negative magnetization $m < 0$. We identified the existence of this phase for  $N=5,6,7,8$, but it could be a finite cluster size effect and should not survive for larger $N$  values. However, it deserves an additional check within other formalisms, like the TBA and TQM methods \cite{Takahashi2002, Campo2015, Angela2020, Sacramento2021, Juttner1998}.

This paper opens up a series of possibilities for future applications. In addition, to a better understanding of the Hubbard model and its application in traditional systems where the model is believed to have a relevant role. As immediate applications of the method, we mention the use of the cluster solutions as a correlated quantum dot (QD), connected to Hubbard chains. This kind of setup can be used to study the Kondo effect \cite{Carolina2021} and different kinds of transport properties. We also studied a simple spintronic application of the CGFM. We study the electronic transport through a 3-site cluster quantum dot (QD) immersed in correlated 3-site Hubbard rectangular conduction bands and obtained spin-polarized currents. The extension of the method to 2D and 3D opens the possibility of applying the method to study high $T_{C}$ superconductivity \cite{Qin2022} and the Mott transition. In applications to 2D or 3D systems, we must change the correlated site's cluster geometry to a closed one, as indicated in Fig. \ref{seed}. Another promising application of the method is in the simulations of ultracold atoms in optical lattices. This research area has defined an ideal platform to verify and explore new physics associated with correlated electronic systems~\cite{Esslinger-AR10, IBloch-NP12, Gross-S17}.

\section*{Acknowledgments}

M.~S.~F. Acknowledges financial support from the Brazilian National Council for Scientific and Technological Development (CNPq) Grant. Nr. 311980/2021-0 and to
Foundation for Support of Research in the State of Rio de Janeiro (FAPERJ) process Nr. 210 355/2018. R.~N~L. Acknowledges financial support from the Brazilian agency Coordination of Improvement of Higher Education Personnel (CAPES) processes Nr. 88882.332178/2019-01 (Ph.D. scholarship) and Nr. 88887.373428/2019-00 (Capes Print program).

\bibliography{References_Hubbard.bib}

\end{document}